\documentclass[epjST,final]{svjour}

\usepackage[utf8]{inputenc}
\usepackage[numbers,sort&compress]{natbib}
\usepackage{amsmath,amssymb,textcomp}
\usepackage[T1]{fontenc}
\usepackage{txfonts}

\usepackage{graphicx}
\usepackage{mdwtab}
\usepackage{url}
\usepackage{algpseudocode,algorithm} 

\newlength{\figwidth}
\graphicspath{{figures/}}

\renewcommand\epsilon{\varepsilon}
\renewcommand\phi{\varphi}
\renewcommand\rho{\varrho}
\renewcommand\theta{\vartheta}
\newcommand\unitvec[1]{\vec{\hat #1}}
\newcommand\tensor[1]{\textsf{\bfseries\slshape #1}}
\newcommand\diff{\mathrm{d}}
\newcommand\grad{\ensuremath{\boldsymbol\nabla}}
\newcommand\divergence{\ensuremath{\boldsymbol\nabla \cdot}}

\newcommand\laplace{\ensuremath{\nabla^2}}

\newcommand\eq[1]{Eq.~\eqref{eq:#1}}
\newcommand\fig[1]{Fig.~\ref{fig:#1}}

\begin{document}

\title{GPU-accelerated simulation of colloidal suspensions with direct
hydrodynamic interactions}
\author{Michael Kopp\inst{1}\fnmsep\inst{2} \and Felix Höf{}ling\inst{1}\fnmsep\inst{2}}

\institute{%
Max-Planck-Institut f\"ur Intelligente Systeme,
Heisenbergstra{\ss}e 3, 70569 Stuttgart
\and
Institut f\"ur Theore\-tische und Angewandte Physik, Universit\"at
Stuttgart, Pfaffenwaldring 57, 70569 Stuttgart, Germany}
\abstract{%
Solvent-mediated hydrodynamic interactions between colloidal particles can
significantly alter their dynamics.
We discuss the implementation of Stokesian dynamics in leading approximation
for streaming processors as provided by the compute unified device architecture (CUDA)
of recent graphics processors (GPUs).
Thereby, the simulation of explicit solvent particles is avoided and
hydrodynamic interactions can easily be accounted for in already available,
highly accelerated molecular dynamics simulations.
Special emphasis is put on efficient memory access and numerical stability.
The algorithm is applied to the periodic sedimentation of a cluster of four
suspended particles.
Finally, we investigate the runtime performance of generic memory access
patterns of complexity $O\bigl (N^2 \bigr)$ for various GPU algorithms relying
on either hardware cache or shared memory.
}

\maketitle

\section{Introduction}

Micron-sized colloidal particles are suspended in a solvent for most
applications.
The particle motion couples to the flow field of the solvent, which can
significantly affect the dynamic properties.
Already the Brownian motion of a single particle in a viscous solvent creates a
slowly decaying flow pattern with feedback on the particle at a later point in
time.
This effect is known as hydrodynamic memory and leads to an experimentally
relevant ``coloured'' noise spectrum~\cite{Franosch:2011}.
The flow field due to a steadily dragged particle, first calculated by George
G.\ Stokes, decays slowly with distance and exerts a drag on nearby particles
referred to as hydrodynamic interactions.
They are intrinsically of many-body nature and in general produce chaotic
trajectories with nontrivial and sometimes surprising effects.
For example, three colloids driven along a circle exhibit an intriguing cyclic
motion~\cite{Lutz:2006} due to the ``peloton effect'': the mobility of a pair of
close colloids is enhanced.
Another example are sedimenting suspensions in a slit pore, where the formation
of complex spatial patterns, related to a Rayleigh--Taylor instability, was
observed in experiments and simulations~\cite{Wysocki:2009, Milinkovic:2011}.
Further, hydrodynamic interactions enhance the self-diffusion of colloidal
particles confined to a fluid interface~\cite{Rinn:1999} and, in the absence of
long-range repulsive forces, accelerate their capillary
collapse~\cite{Bleibel:2011}.

Computer simulations of colloidal suspensions including hydrodynamic effects are
challenged by the vast amount of solvent molecules per colloid particle.
Various techniques have been developed which use coarse-grained solvents to
approximate the fluid properties at mesoscopic scales.
Popular methods are the lattice-Boltzmann technique and multi-particle collision
dynamics, see, e.g., Refs.~\citealp{Duenweg:2009, Gompper:2009, Padding:2006}
and references therein.
In particular for diluted suspensions, such approaches have the drawback that
the fluid representation consumes most of the computational resources.
Complementary to methods with explicit solvent are Stokesian dynamics
simulations~\cite{Ermak:1978, Durlofsky:1987}, where the
solvent-mediated interactions are accounted for by a many-particle mobility
matrix constructed from hydrodynamic tensors leading to deterministic equations
of motion.
The correct incorporation of Brownian motion~\cite{Ermak:1978} is
computationally expensive, and often approximate algorithms are
used~\cite{Banchio:2003}.
Here, we restrict to phenomena where Brownian motion is of relatively low
importance compared to the overall motion (large Péclet number) and can be
neglected, examples can be found in Refs.~\citealp{Bleibel:2011} and
\citealp{Ladd:1993, Padding:2004, Putz:2010}.

Computing hardware has seen a paradigm shift during the last decade from
single-core processors to highly parallel multi-core architectures.
Driven by the consumer market for video games, graphics processing units (GPUs)
have been developed that contain several hundred tiny processor cores on a
single chip.
These subunits are specialised on streaming numerical computations of large
data sets in parallel.
Today, such streaming processors are often part of new installations in
high-performance computing centres.

The success of conventional molecular dynamics simulations on
GPUs~\cite{Baker:2011} suggests to exploit the potential of GPUs also for
simulation techniques with more complex interactions and advanced algorithms.
In general, particle-based continuum simulations are more challenging than
lattice models with respect to algorithms and runtime
performance~\cite{Anderson:2008, Meel:2008} as well as numerical stability and
floating-point precision~\cite{Glassy_GPU:2011, Ruymgaart:2011}.
Here, we will specifically address simulations of suspended particles with
direct hydrodynamic interactions instantaneously mediated by an implicit
solvent.
GPUs were used previously to simulate hydrodynamic interactions in suspensions
of active dumbbells~\cite{Putz:2010}.
A hybrid implementation scheme with explicit solvent is described in
Ref.~\citealp{Roehm:2012}, where the particles are treated by the host
processor(s) and the solvent is modelled as lattice-Boltzmann fluid and
propagated by the GPU.

\section{Theoretical background}
\label{sec:theo}

\subsection{Stokes equation}

The starting point for a mesoscopic description of the solvent is the
Navier--Stokes equation for incompressible flow (low Mach number) of Newtonian
fluids~\cite{Batchelor:FluidDynamics},
\begin{equation}
  \rho \left( \partial_t \vec u + \vec u \cdot \grad \vec u \right)
  =
  \eta \laplace \vec u - \grad  p + \vec f \,,
  \qquad \divergence \vec u = 0 \,,
  \label{eq:navier-stokes}
\end{equation}
$\vec u = \vec u(\vec r, t)$ denotes the velocity field of the fluid, $\rho$ the
uniform mass density, $\eta$ the shear viscosity, $p=p(\vec r, t)$ the pressure
field, and $\vec f=\vec f(\vec r, t)$ an external force density.
Such a continuum approach holds at length and time scales much larger than the
molecular scales of the solvent (low Knudsen number).
The assumption of incompressibility is quite well fulfilled by most fluids at
standard temperature and pressure.

Suspended colloidal particles impose additional boundary conditions on the
fluid.
For not too small particles, there is ample experimental evidence that the fluid
molecules at the particle boundary adopt the velocity of the latter, which is
referred to as stick boundary conditions,
\begin{equation}
  \vec u(\vec r) = \vec v  + \vec \Omega \times (\vec r - \vec r_0)
  \quad \text{for} \quad \vec r \in \partial V.
  \label{eq:stickBoundary}
\end{equation}
Here, $\vec v$ and $\vec \Omega$ are the translational and angular velocities of
the immersed object, respectively; $\vec r_0$ is a reference point of the
particle and $\partial V$ denotes its surface.

The physics of micron-sized objects like bacteria or colloidal particles is
mostly dominated by viscous friction (low Reynolds number), which permits
neglecting the inertia term $\rho \vec u \cdot\grad \vec u$  in
\eq{navier-stokes}, known as creeping flow limit.
The time scale $\tau$ for the velocity field to adjust to a local disturbance is
set by the diffusive propagation of shear waves together with the largest
typical length scale $L$ of the problem~\cite{Dhont:ColloidDynamics,
Eckhardt:2008}; for water and $L=10$\,\textmu{}m,
$\tau \approx \rho L^2/\eta \approx 10^{-4}$\,s.
For sufficiently slowly changing boundary conditions, the fluid response can be
considered instantaneous and the time derivative $\rho \partial_t \vec u$ can be
dropped as well, which leads to the time-independent Stokes equation,
\begin{equation}
   - \eta \laplace \vec u = -\grad p + \vec f, \qquad \divergence \vec u = 0 \,,
  \label{eq:stokes}
\end{equation}
which forms the basis of Stokesian dynamics simulations. It has some analogies
to the Poisson equation in magnetostatics, from which many concepts can be
carried over.
The approximation of infinitely fast propagating hydrodynamic interactions is a
serious limitation of Stokesian dynamics and needs to be justified for every
problem, in particular when considering large system sizes or an externally
prescribed time-dependence of the boundaries~\cite{Eckhardt:2008}.

\subsection{Hydrodynamic interactions}

The flow field $\vec u(\vec r) = \tensor G(\vec r) \cdot \vec F$ due to a point
force, $\vec f(\vec r) = \vec F \, \delta(\vec r)$, acting on the fluid defines
Green's function of the Stokes equation, the Oseen
tensor~\cite{Oseen:Hydrodynamik},
\begin{equation}
  \tensor G(\vec r) = \frac{1}{8 \pi \eta r} (\tensor I +  \unitvec r \unitvec r) \,;
  \label{eq:oseenTensor}
\end{equation}
the notation $\vec a\vec b$ indicates a second-rank tensor with components
$(\vec a \vec b)_{\alpha\beta} = a_\alpha b_\beta$,
$\tensor I_{\alpha\beta}=\delta_{\alpha\beta}$ is the unit tensor, and
$\unitvec r = \vec r / r$ with $r=|\vec r|$.
The most prominent property of the Oseen tensor is the slow $1/r$-decay of its
magnitude reflecting the long-range nature of hydrodynamic interactions.

The boundary condition, \eq{stickBoundary}, for the suspended particles can be
replaced by an (\emph{a priori} unknown) induced force density $\vec f(\vec r)$
localised on the particle surfaces, continuing the velocity field inside the
particles.
By the Stokes equation \eqref{eq:stokes}, $\vec f(\vec r)$ generates a flow
field $\vec u(\vec r)$, which in general depends on the shapes and orientations
of the particles.
The contribution $\vec u_0(\vec r)$ from a single particle in a quiescent fluid
is obtained within a multipole expansion analogously to the treatment in
magnetostatics~\cite{Mazur:1982}.
Keeping only the leading, monopole contribution, it reads
\begin{equation}
  \vec u_0(\vec r)= \int\! \diff^3  r'\, \tensor G(\vec r - \vec r') \cdot \vec f(\vec r')
  = \left(1 + \frac{a^2}{6} \laplace \right)  \tensor G(\vec r) \cdot \vec F
  + O\bigl(r^{-4}\bigr) \quad \text{for} \quad r \gg a,
  \label{eq:greens}
\end{equation}
with total force $\vec F = \int \! \vec f(\vec r) \, \diff^3  r$,
(root-mean-square) radius of the particle surface $a$, and the particle being
centred at the coordinate origin.
An external drag force can be included in the force monopole, and the flow field
for many particles is found by superposition of the single particle
contributions.

The velocity distribution on the surface of another particle at position $\vec
r$ may be expanded in multipoles as well, the monopole term yields just the
particle's translational velocity $\vec v$; its computation is facilitated by
Faxén's theorem~\cite{Dhont:ColloidDynamics},
\begin{equation}
  \vec v = \frac{1}{4\pi a^2} \int_{\partial V(\vec r)}\! \vec u(\vec r') \,\diff^3 r'
  = \left(1 + \frac{a^2}{6} \laplace \right)  \vec u(\vec r).
  \label{eq:faxen}
\end{equation}
The velocity dipole determines the angular velocity, and higher terms vanish for
stick boundary conditions.
Rotational motion leads to a coupling between velocity and force monopoles,
which is of order $r^{-4}$ and, at the monopole level, limits the consistent
expansion of \eq{greens} to the third order in distance.

Combining Eqs.~\eqref{eq:greens} and \eqref{eq:faxen} for $N$ particles encodes
the hydrodynamic interactions in a configuration-dependent many-body mobility
matrix $\tensor \textmu_{ij}(\vec r_1, \dots, \vec r_N)$ relating forces to
velocities,
\begin{subequations}
\label{eq:rotneMobility}
\begin{equation}
  \vec v_i = \sum_{j =1}^N \tensor \textmu_{ij} \cdot \vec F_j
  = \mu_0 \vec F_i +
  \sum_{j \neq i}^N \tensor T(\vec r_i - \vec r_j) \cdot \vec F_j
  \quad \text{for} \quad i=1,\dots, N;
\end{equation}
Latin subscripts refer to particle indices.
The self-contribution ${\tensor \textmu}_{ii}$ describes the direct effect of a
drag force on the particle itself and is set by the Stokes mobility,
$\mu_0=1/6\pi\eta a$.
The off-diagonal terms are given by the Rotne--Prager tensor,
\begin{equation}
  \tensor T(\vec r) = \left(1 + \frac{a^2}{3} \laplace \right) \tensor G(\vec r)
  = \mu_0 \frac{3 a}{4 r} (\tensor I + \unitvec r \unitvec r)
    + \mu_0 \frac{a^3}{2 r^3} (\tensor I - 3 \unitvec r \unitvec r) \,.
\end{equation}
\end{subequations}
If the particle size is negligible compared to the particle separation, the
gradients in Eqs.~\eqref{eq:greens} and \eqref{eq:faxen} may be dropped,
replacing $\tensor T$ by the Oseen tensor $\tensor G$.  An advantage of
the Rotne--Prager tensor, necessary for the simulation of Brownian motion, is
that the matrix $\tensor \textmu$ is positive definite if no pair of particle
centres is closer than $2a$~\cite{Rotne:1969}.

\subsection{Lubrication correction}

If two particles come close together, the hydrodynamic coupling is only poorly
described by the far-field expansion.
In particular, the many-body friction matrix, $\boldsymbol \zeta = \tensor
\textmu^{-1}$, becomes singular if only a thin lubrication film separates the
particle surfaces.
\citet{Cox:1974} computed the leading singularities for two approaching surfaces
of arbitrary shape, and asymptotic expansions for two spheres were given by
\citet{Jeffrey:1984}.
Let us introduce relative positions and velocities, $\vec r_{ij}=\vec r_i - \vec
r_j$ and  $\vec v_{ij}=\vec v_i - \vec v_j$, and the dimensionless separation
between particle surfaces, $s_{ij}=r_{ij}/a - 2$.
Then, the singular part of the force on sphere $i$ due to a sphere $j$ moving
nearby reads~\cite{Jeffrey:1984}
\begin{equation}
  \mu_0 \vec F^\text{sing}_i =
    \left[ \frac{1}{4s_{ij}} - \frac{9}{40}\log(s_{ij}) \right]
    \unitvec r_{ij} \unitvec r_{ij} \cdot \vec v_{ij}
    - \frac{1}{6} \log(s_{ij}) \, (\tensor I-\unitvec r_{ij} \unitvec r_{ij}) \cdot \vec v_{ij}
    + O\bigl(s_{ij}^0\bigr) \,,
  \label{eq:lubrication}
\end{equation}
again ignoring contributions from rotational motion.
We truncate \emph{ad hoc} at $s_{ij}=1$ or $r_{ij}=3a$, where the logarithm
changes its sign, avoiding a sign change of the lubrication forces in the present
approximation.
The lubrication correction to the many-body friction matrix is constructed from
close particle pairs alone, such that
$\vec F^\text{sing}_i = \sum_j \boldsymbol \zeta^\text{sing}_{ij} \vec v_j$.
The result from lubrication theory has to be matched with the far-field
expansion of the inverse mobility matrix~\cite{Durlofsky:1987, Ladd:1990}.
Since we kept only the singular terms of $\boldsymbol \zeta$ in
\eq{lubrication} it can simply be added,
\begin{equation}
  \tensor v
  = \left(\tensor \textmu^{-1} + \boldsymbol \zeta^\text{sing} \right)^{-1} \tensor F
  = \tensor \textmu
    \left(\tensor I + \boldsymbol \zeta^\text{sing} \tensor \textmu \right)^{-1} \tensor F \,.
  \label{eq:mobilityCorrection}
\end{equation}

\section{Simulation algorithm and implementation}
\label{sec:impl}

\subsection{Algorithm}

The idea behind Stokesian dynamics simulations is to avoid the explicit
simulation of the individual solvent molecules as one would do in traditional
molecular dynamics simulations.
Rather, hydrodynamic interactions between colloidal particles are properly
incorporated by means of the many-body mobility matrix $\tensor \textmu$ defined
in \eq{rotneMobility}.
This matrix is entirely constructed from pair contributions, even at higher
orders of the multipole expansion.

Our main motivation for including the lubrication correction is the
implementation of a hard core repulsion. Hence, we suggest the approximation
$\tensor v \approx \tensor \textmu
\bigl(\tensor I - \boldsymbol \zeta^\text{sing} \tensor \textmu \bigr)
\tensor F$ to \eq{mobilityCorrection}, which avoids the storage and inversion of
the $3N\times 3N$ matrix~$\tensor \textmu$.
Further, we approximate on the right-hand side
$\tensor \textmu \boldsymbol \zeta^\text{sing} \tensor \textmu \tensor F
\approx \tensor \textmu \boldsymbol \zeta^\text{sing} \tensor v$
using the velocities from the previous simulation step; effectively, the
particle forces are corrected for lubrication before the velocity update.
Then, the simulation algorithm consists of repeating the following steps:
\begin{enumerate}
  \item update particle positions: \\ $\vec r_i \leftarrow \vec r_i + \delta t\, \vec v_i$
  \item compute forces on each particle due to external forces, interactions,
    and lubrication: \\
  $
    \tensor F_i \leftarrow \vec F_i^\text{ext}(\vec r_i)
      + \sum_j ' \vec F_i^\text{int}(\vec r_i, \vec r_j)
      - \sum_j ' \boldsymbol \zeta^\text{sing}_{ij}(\vec r_i, \vec r_j) \cdot (\vec v_i - \vec v_j)
  $
  \item compute particle velocities via mobility matrix: \\
  $\vec v_i \leftarrow \sum_j \tensor \textmu_{ij}(\vec r_i, \vec r_j) \cdot \vec F_j$
\end{enumerate}
Particle positions are propagated by a simple Euler integration in step~1.
The primed sums in step~2 can be accelerated by Verlet neighbour lists common
for short-ranged forces as described, e.g., in Ref.~\citealp{Glassy_GPU:2011}.
Since the hydrodynamic interactions are long-ranged, like electrostatics or
gravity, it is preferable to keep all interactions of every pair of particles
and not to cut off at some short distance.\footnote{%
We use a simulation box with periodic boundaries and include only hydrodynamic
interactions from the nearest image of each particle. Then, the finite size of
the box effectively imposes a cutoff at large distances. Having applications to
colloidal dispersions in mind, the long-ranged interaction with all periodic
images is ignored since the periodicity has no physical grounds and is as
artificial as a sharp cutoff.}
Thus, the sum in step~3 is the computationally most intense part involving $N^2$
contributions for all $N$ particles; the mobility matrix itself, however, need
not be stored in memory.


\subsection{GPU implementation}
\label{sec:gpu_impl}

The present work relies on the ``compute unified device architecture'' (CUDA)
introduced by the Nvidia Corporation~\cite{CUDA_Guide}.
It provides some abstraction to the actual GPU hardware and makes Nvidia GPUs
freely programmable using the C++ language.
The central concept is thread parallelism, which means that several hundred or
thousand program threads execute the same routine on the GPU, the latter is
referred to as ``kernel''.
Each group of (currently) $M=32$ threads forms a ``warp'', several warps are
organised into a thread block, and all blocks together represent the full set
of threads.
The threads within a warp execute simultaneously, but the execution order of
warps is not guaranteed.
The warps comprising a block are resident on the same multiprocessor and
execute independently, the size of a block is limited by the available hardware
resources.
Communication between threads is only supported within the same block using
``shared'' memory, which does not persist beyond the lifetime of the block.
Data input and output of a kernel occurs through the ``global'' memory, which is
the only persistent type of memory on the GPU board.
Despite the high overall bandwidth of global memory, each memory transaction
entails a huge latency of several hundred clock cycles, which is, however,
partly hidden by the thread scheduler on the chip.
Optimal access to global memory is ensured if a warp of threads reads from or
writes to a sequence of contiguous memory locations that fit within a single,
properly aligned window of 64 or 128 bytes.
Then, the memory access of the whole warp is coalesced into a single transaction
of a large block.
As another remedy for the high memory latency, the Nvidia GPUs of CUDA compute
capability $\geq 2.0$ (``Fermi'') have been equipped with hardware caches.
We put particular attention to memory transfer because it is typically the
bottleneck of a GPU program; it appears to become the limiting factor in
conventional high-performance computing as well.
We are not concerned with memory transfer between host system and GPU device
since our simulation algorithms are fully implemented for the GPU.


\begin{algorithm}[tb]
  \caption{Pseudo code of the CUDA kernel for updating the particle velocities in
  the \emph{na{\"i}ve} implementation, for an illustration see
  \fig{implementations}a.  The kernel is executed concurrently by at least $N$
  threads, array indices are 1-based, and arrays in global memory are denoted by
  capital letters.}
  \label{algo:naiveKernel}
\begin{algorithmic}[1]
  \Function{update\_velocity}{position array $\vec R$,
    velocity array $\vec V$, force array $\vec F$, \#particles $N$, \dots}
    \State local $j \gets \text{global thread ID}$
    \State local $\vec r \gets  \vec R_j$ \Comment{read from global memory}
    \State local $\vec v \gets 0$
    \For{$k$ from 1 to $N$} \Comment{loop over all particles}
      \State local $\vec r', \vec f' \gets \vec R_k, \vec F_k$ \Comment{read from global memory}
      \label{algo:naiveKernel:readglobal}
      \If{$j = k$}\label{algo:naiveKernel:ifself}
        \Comment{velocity contribution according to \eq{rotneMobility}}
        \State $\vec v \gets \vec v + \mu_0 \, \vec f'$. \Comment{self-contribution}
      \Else
        \State $\vec v \gets \vec v + \tensor T(\vec r - \vec r') \cdot \vec f'$.
          \Comment{use hydrodynamic tensor}
      \EndIf
    \EndFor
    \State $\vec V_j \gets \vec v$ \Comment{write to global memory}
  \EndFunction
\end{algorithmic}
\end{algorithm}

\begin{figure}
  \centering \tabcolsep=0pt
  \begin{tabular}{p{.48\linewidth}@{\hspace{.04\linewidth}}p{.48\linewidth}}
  \textbf{a} & \textbf{b} \\[-1.5\baselineskip]
  \vspace{0pt}\includegraphics[width=\linewidth]{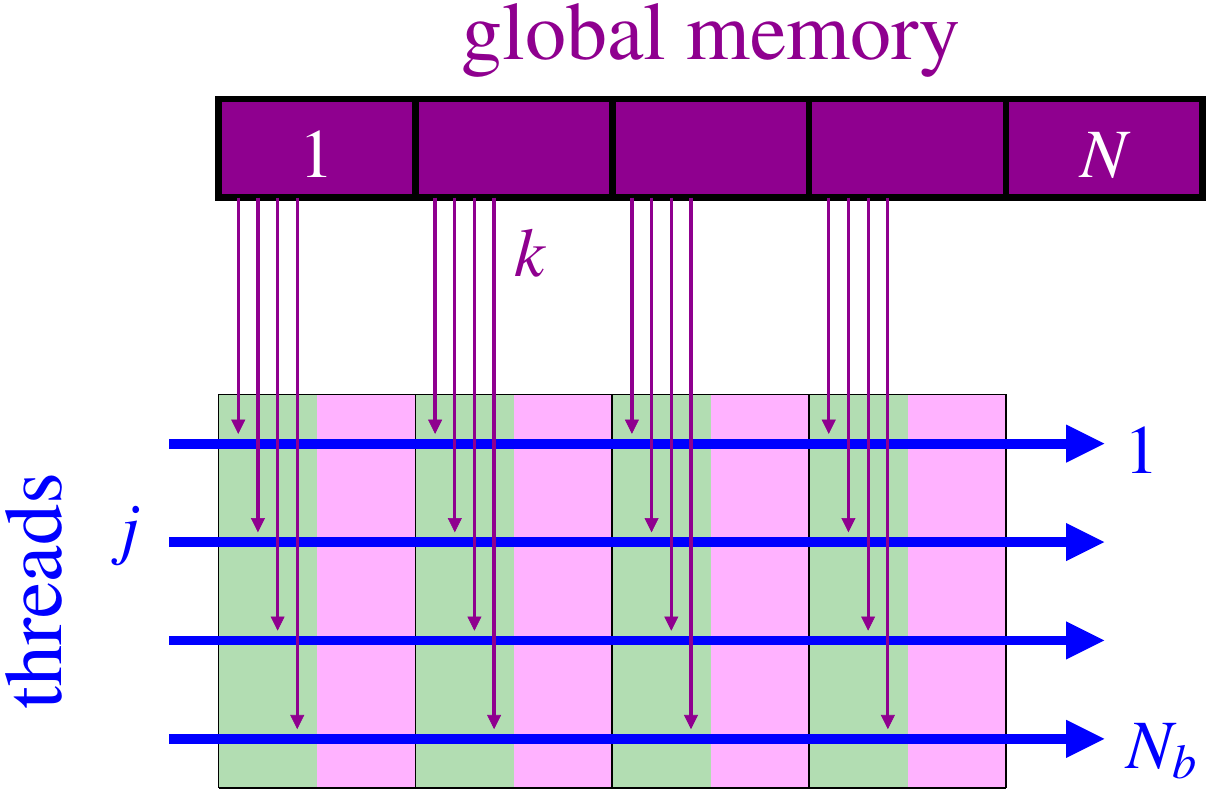} &
  \vspace{0pt}\includegraphics[width=\linewidth]{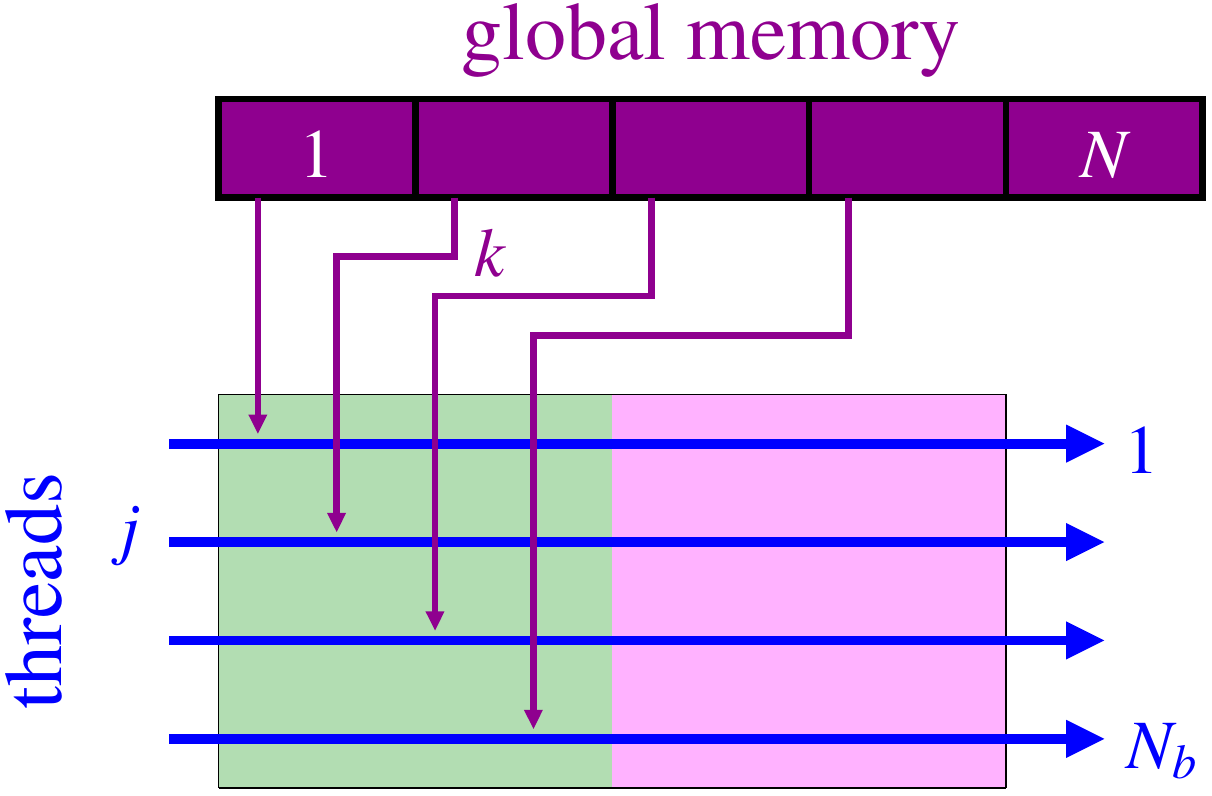} \\[\baselineskip]
  \textbf{c} & \textbf{d} \\[-1.5\baselineskip]
  \vspace{0pt}\includegraphics[width=\linewidth]{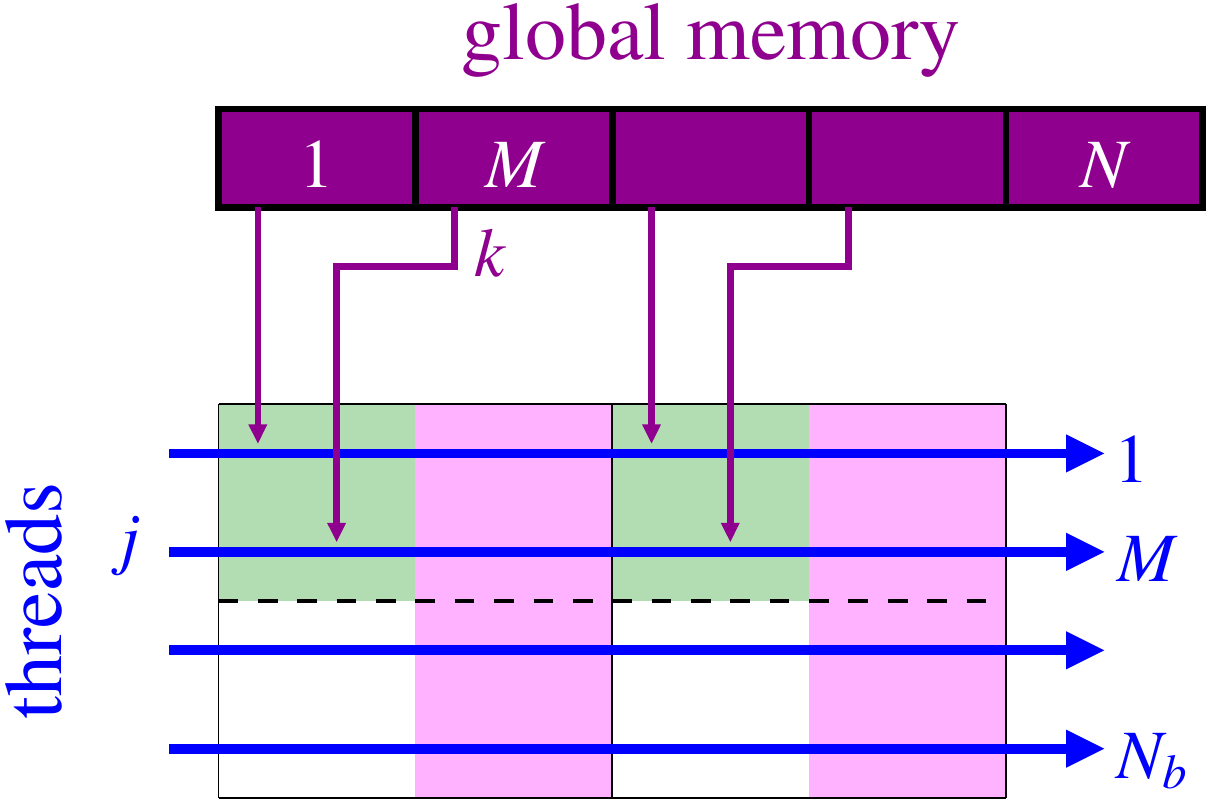} &
  \vspace{0pt}\hfill\includegraphics[width=\linewidth]{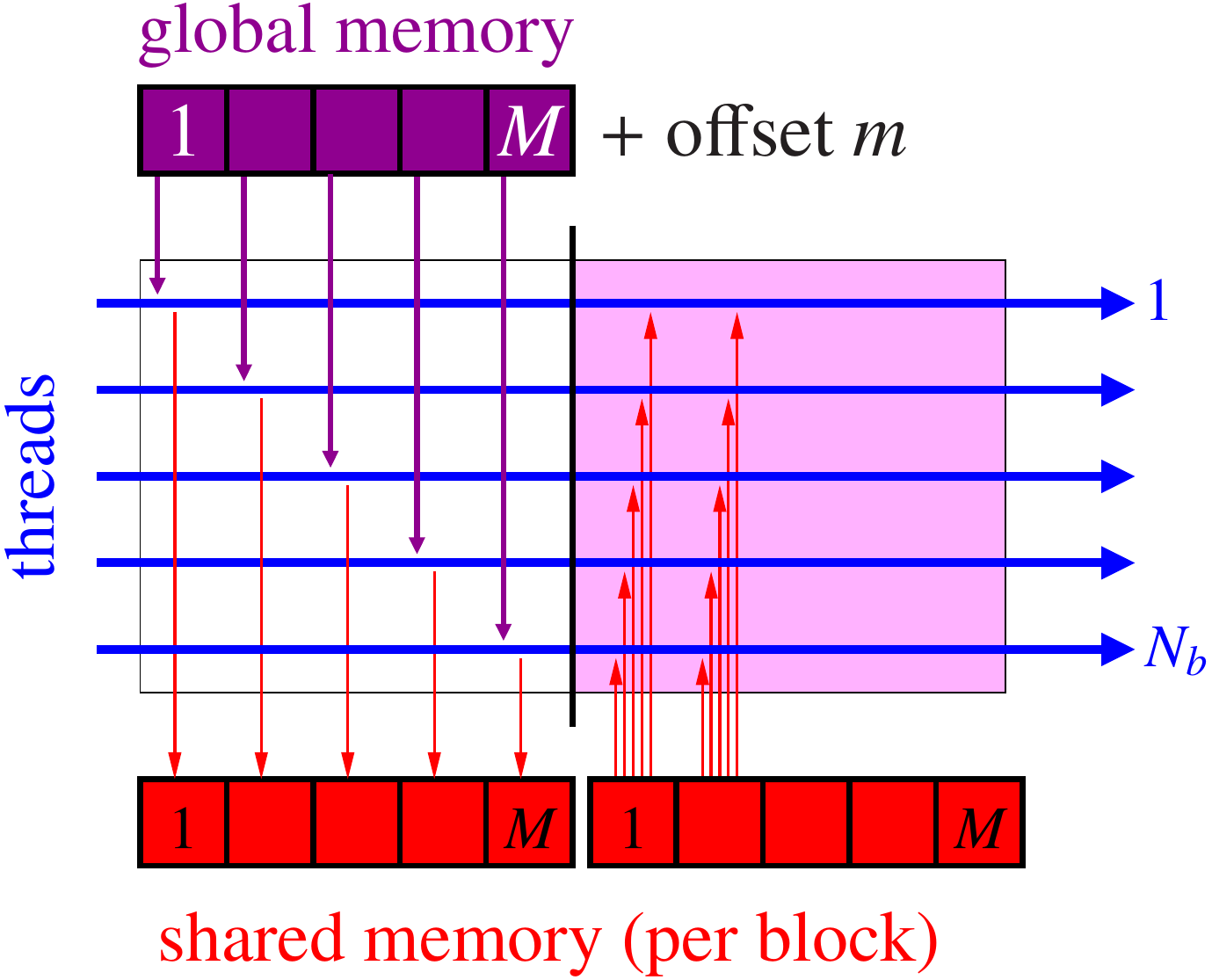}\hfill
  \end{tabular}
  \caption{Visualisation of various algorithms for the computation of
  hydrodynamic interactions, see Algorithms~\ref{algo:naiveKernel} and
  \ref{algo:smalltilingKernel}.
  Blue horizontal arrows represent the threads being associated with one
  particle each, gradually computing hydrodynamic interactions with all $N$
  particles (purple boxes on top); only one block of $N_b$ threads is shown.
  Purple, vertical arrows represent global memory accesses; green rectangles
  symbolise data fetching and pink ones data processing, i.e., the actual
  computation.
  ~(a) na{\"i}ve implementation. Data of one particle at a time are read from
  global memory directly without caching.
  ~(b) tiling algorithm. Data are read from global memory by coalesced
  transactions and are stored blockwise in shared memory for further processing.
  ~(c) small-tiling algorithm using small tiles of size $M$, it is similar to
  the tiling algorithm, but the coalesced read is performed only by one warp of
  each thread block (delimited by the dashed line).
  ~(d) access to shared memory (red arrows/box) within one thread block for the
  tiling ($M=N_b$) and small-tiling algorithms.
  }
  \label{fig:implementations}
\end{figure}

The most straightforward implementation of step~3 on a GPU is to assign one
thread to each particle, which computes and accumulates the contributions from
all $N$ particles (including the self-contribution) to the velocity of the
thread's particle, see Algorithm~\ref{algo:naiveKernel}.
The symmetry $\tensor \textmu_{jk} = \tensor \textmu_{jk}^T$ will not be made
use of for the same reason why Newton's third law is not exploited for the
computation of forces from pair interactions: transferring the result from
thread~$j$ to an arbitrary thread~$k$ would break thread parallelism and would
require a sophisticated communication pattern involving a bloat of global memory
transactions.
The kernel basically consists of a loop over \emph{all} particles, sequentially
reading and processing their positions and forces from global memory.
Before exit, the kernel stores the computed velocity, which is the only global
write operation.
The threads within a warp can diverge (line~\ref{algo:naiveKernel:ifself}) due
to the self-contribution of a particle to its velocity, which unavoidably
happens once for every thread in a warp and does not seem to significantly
impact the execution time.

\begin{algorithm}[tb]
  \caption{The \emph{small-tiling} variant of algorithm~\ref{algo:naiveKernel}.
  Position and force data are partitioned in small tiles and prefetched into
  shared memory by the first warp of each block, for an illustration see
  \fig{implementations}c,d. Arrays in shared memory are denoted by a tilde.}
  \label{algo:smalltilingKernel}
\begin{algorithmic}[1]
  \Function{update\_velocity}{position array $\vec R$,
    velocity array $\vec V$, force array $\vec F$, \#particles $N$, \dots}
    \State local $j \gets \text{global thread ID}$
    \State local $n \gets \text{thread ID within block}$
    \State local $M \gets \text{warp size}$
    \State allocate shared memory $\tilde{\vec r}[M], \tilde{\vec f}[M]$
      \Comment{positions, forces}
    \State local $\vec r \gets  \vec R_j$ \Comment{read from global memory}
    \State local $\vec v \gets 0$
    \For{$m$ from $1$ to $N$ with stride
    $M$}\label{algo:smalltilingKernel:strideloop}
      \Comment{loop over all particles in tiles of size $M$}
      \If{thread in first warp}\label{algo:smalltilingKernel:ifsmalltiling}
        \State $\tilde{\vec r}_n, \tilde{\vec f}_n \gets \vec R_{m + n}, \vec
        F_{m + n}$ \label{algo:smalltilingKernel:readglobal}
          \Comment{copy from global to shared memory}
      \EndIf
      \State synchronise threads \label{algo:smalltilingKernel:syncread}
        \Comment{wait until data are fetched}
      \Statex

      \For{$k$ from $1$ to $M$}
        \Comment{loop over fetched data}
        \State local $\vec r', \vec f' \gets \tilde{\vec r}_{k}, \tilde{\vec f}_{k}$
          \Comment{read from shared memory}
        \If{$j = m + k$}\label{algo:smalltilingKernel:ifself}
          \Comment{velocity contribution according to \eq{rotneMobility}}
          \State $\vec v \gets \vec v + \mu_0 \, \vec f'$.  \Comment{self-contribution}
        \Else
          \State $\vec v \gets \vec v + \tensor G(\vec r - \vec r') \cdot \vec f'$.
          \Comment{use hydrodynamic tensor}
          \label{algo:smalltilingKernel:accumulate}
        \EndIf
      \EndFor
      \State synchronise threads \label{algo:smalltilingKernel:synccompute}
        \Comment{wait until data are processed}
    \EndFor
    \State $\vec V_j \gets \vec v$ \Comment{write to global memory}
  \EndFunction
\end{algorithmic}
\end{algorithm}

Following \citet{Nyland:2007}, the na\"ive read pattern in
line~\ref{algo:naiveKernel:readglobal} is excessively inefficient: first, the
same location in memory is accessed by every thread (\fig{implementations}a)
and, second, memory bandwidth is wasted.
The requirements for optimal memory access can be met by tiling the matrix of
$N\times N$ interactions into squares of size~$N_b$~\cite{Nyland:2007}, see
\fig{implementations}b; a similar approach has been proposed for general matrix
multiplication~\cite{CUDA_Guide}.
$N_b$ denotes the number of threads per block, it varies typically between 128
and 1024.
The interactions are processed tile by tile and the data for a tile are cached
in shared memory, making them available to all threads within the block
(\fig{implementations}d), and reducing the number of global memory reads by a
factor of~$N_b /M$.
For gravity~\cite{Nyland:2007} or matrix multiplication~\cite{CUDA_Guide}, the
size of a data item is smaller than in our case of hydrodynamic interactions.
Here, the information per particle comprise two three-dimensional vectors, a
position and a force, amounting to 32 bytes per particle
(2\,\texttimes\,\textsf{float4}) if alignment restrictions are obeyed.
The relatively high shared memory usage of $32\times M=1024$ bytes per warp
limits the multiprocessor occupancy on older hardware (CUDA compute capability <
2.0) with a potential increase of kernel execution time; for compute
capability 2.0, maximum occupancy is exhausted by 1\,kb of shared memory per
warp \cite{CUDA_Guide}.
The shared memory usage can be significantly reduced without increasing the
volume of global memory transfer if merely the first warp of a block fetches the
data, the tiles become of rectangular shape $M \times N_b$.
The resulting \emph{small-tiling} version is displayed in
Algorithm~\ref{algo:smalltilingKernel} and illustrated in
\fig{implementations}c.
The original \emph{tiling} algorithm is recovered by replacing $M$ with the
block size and removing the condition in
line~\ref{algo:smalltilingKernel:ifsmalltiling}.

Data fetching from global memory uses adjacent memory locations and allows for
coalescable access (line~\ref{algo:smalltilingKernel:readglobal}), in contrast
to line~\ref{algo:naiveKernel:readglobal} of Algorithm~\ref{algo:naiveKernel}.
It is followed by a synchronisation barrier
(line~\ref{algo:smalltilingKernel:syncread}) to prevent the other warps from
premature reading from shared memory.
The condition in line~\ref{algo:smalltilingKernel:ifsmalltiling} does not break
convergent thread execution within each warp (the condition is fulfilled
either by each thread of a warp or by none),
and only memory access within a warp is eligible for coalesced transactions.
We see only the slight disadvantage with respect to execution time that fewer
transactions are pending on a multiprocessor, which impedes latency hiding by
the scheduler.
After data fetching is completed, all threads of the block read from shared
memory and compute the velocity contribution from hydrodynamic interactions for
their respective particle.
The tile is closed by another synchronisation barrier
(line~\ref{algo:smalltilingKernel:synccompute}) signalling that the cached data
may be overwritten.

The velocity sum in \eq{rotneMobility} contains contributions of  very different
magnitude since the terms decay like the inverse particle separation,
$1/r_{ij}$.
Conveying the experience from conventional MD
simulations~\cite{Glassy_GPU:2011}, the numerical error can significantly be
reduced if the accumulation in line~\ref{algo:smalltilingKernel:accumulate} of
Algorithm~\ref{algo:smalltilingKernel} is performed with higher floating-point
precision, whereas the hydrodynamic tensor and the forces are computed in
single, 24-bit precision.
Still reading particle positions and forces in single precision keeps down the
memory transfer volume.
For CUDA compute capability $\geq 2.0$, native double precision may be used, for
older hardware, an emulation via double-single arithmetic is available and can
even be more efficient~\cite{Glassy_GPU:2011}.
Finally, since an all-pairs interaction does not involve selective reading,
e.g., via Verlet neighbour lists, the physical positions of the particles do not
affect the kernel execution time and particle reordering is not necessary.
It is, however, required for the efficient computation of interparticle forces
and the lubrication correction.

\subsection{Runtime performance}

\begin{figure}
  \centering
  \centering \tabcolsep=0pt
  \begin{tabular}{m{.6\linewidth}m{.4\linewidth}}
  \textbf{a} & \textbf{b} \\[-\baselineskip]
  \includegraphics[height=.67\linewidth]{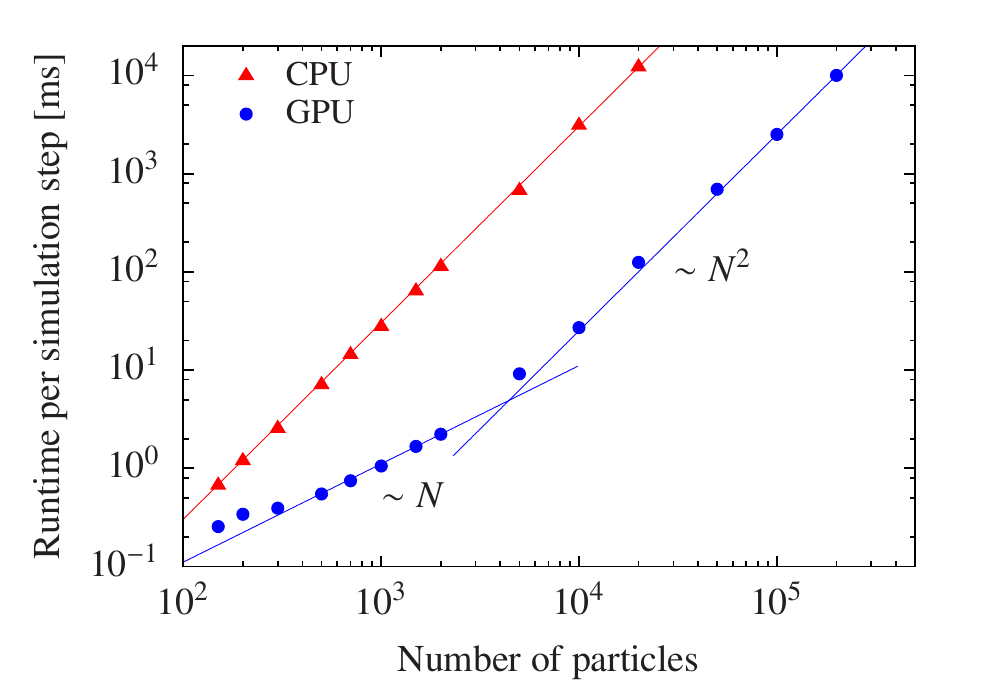} &
  \includegraphics[height=.99\linewidth]{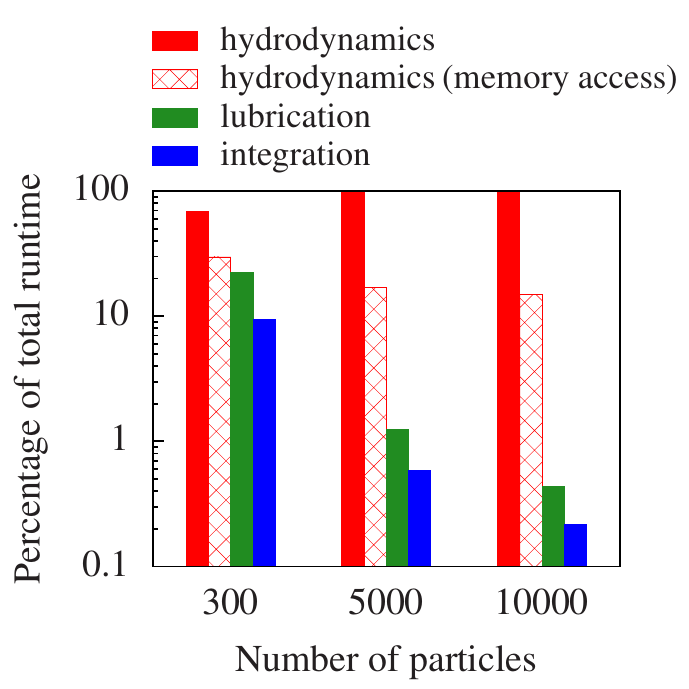}
  \end{tabular}
  \caption{(a) Simulation runtime per step for the computation of hydrodynamic
  interactions using CPU and GPU implementations of
  Algorithm~\ref{algo:smalltilingKernel};  straight lines are fits to the
  regimes of linear and quadratic scaling with particle number.
  ~(b) Relative contributions to the total GPU runtime from the computation of
  hydrodynamic interactions (hatched bars memory access only, without
  computation), of lubrication forces, and integration of the equations of
  motion; the chosen particle numbers correspond to the two scaling regimes and
  the crossover region.
  }
\label{fig:cpugpu}
\end{figure}

The described scheme for the computation of hydrodynamic interactions,
Algorithm~\ref{algo:smalltilingKernel}, has been implemented as a module extending
\emph{HAL's MD package}~\cite{HALMD}, which forms the basis for the following
measurements of runtime performance.
The benchmarks mimic the simulation of a sedimenting suspension: all particles
experience the same constant drag force $\vec F=(0,0,-F)$ due to their buoyant
weight, which together with the particle radius~$a$ and the Stokes mobility
$\mu_0$ defines the unit of time, $\tau_s=a/\mu_0 F$.
The particles were initially placed on an fcc lattice with a fixed number
density of $n=N/L^3=0.1 a^{-3}$, and the number of particles $N$ was varied
between 150 and 200,000.
Particle positions were propagated by the Euler integrator using a timestep of
$0.001 \tau_s$, and simulations were run for 100 steps.
The GPU hardware for the benchmarks was an Nvidia Tesla C2050, the program was
compiled with the CUDA software development kit (SDK) version 4.0 using the
option \textsf{--arch sm\_13}, and kernels were launched with 512 threads per
block.
Note that the simulations were about 2.3-times slower on the same hardware if
the compiler optimised for the actual target architecture with \textsf{--arch
sm\_20}, probably due to a different implementation of floating-point
arithmetic.\footnote{%
The less IEEE-compliant floating-point arithmetic of CUDA compute capability
1.3 is approximately restored by the additional CUDA compiler flags
\textsf{--ftz=true --prec--div=false --prec--sqrt=false}.}
For reference, we compare with an equivalent serial implementation for the CPU
(Intel Xeon E5620 clocked at 2.40\,GHz), with the only difference that the
symmetry $\tensor \textmu_{ij}=\tensor \textmu_{ji}$ is exploited.

The measured runtimes per simulation step are displayed in \fig{cpugpu}.
One infers clearly that the runtime of the GPU implementation scales
quadratically in $N$ for large system sizes, while it grows only linearly for $N
\lesssim 2,000$ particles.
The former quadratic scaling is also found for the CPU version for all particle
numbers, simply reflecting that the interaction between all particle pairs
requires $O\bigl(N^2\bigr)$ computations.
In the scaling regime, the speedup factor of the GPU over the CPU is about 120.
The linear scaling for not too large particle numbers originates from a partial
usage of GPU resources: doubling the number of particles, the execution time of
each CUDA thread doubles, but so does the number of threads running concurrently.
In summary, the total kernel runtime only doubles.
When the device capacity is exhausted not all threads can run in parallel
anymore, yielding the quadratic increase of the runtime with the number of
particles.
A small performance gain of 13\% is achieved by computing the hydrodynamic
interactions only to order $O\bigl(r^{-1}\bigr)$, i.e., in the Oseen approximation.

In Section~\ref{sec:gpu_impl}, we developed the small-tiling algorithm for
efficient access to global memory.
Surprisingly, we found that a variant of the eventual kernel for the computation
of hydrodynamic interactions using the na{\"i}ve memory access pattern slightly
\emph{outperforms} the more sophisticated algorithm, the runtime in the scaling
regime is reduced by 8\%.
This observation motivated us to scrutinise the performance of the different
memory access algorithms depending on the specific access pattern, see
Appendix~\ref{sec:compAlgo}.
For linear data access, the na\"ive implementation performs increasingly better
as the size of the individual data elements increases, and finally the overhead
of data caching penalises the tiling algorithms.
In the present case, both criteria are met: particle data are laid out linearly
and 32~bytes are read for each interaction.

\section{Numerical stability: periodic sedimentation of four spheres}

\begin{figure}
    \begin{center}
    \tabcolsep=0pt
    \begin{tabular}{m{.4\linewidth}m{.6\linewidth}}
        \textbf{a} & \textbf{b} \\[-\baselineskip]
        \includegraphics[width=\linewidth, trim=15 15 15 15]{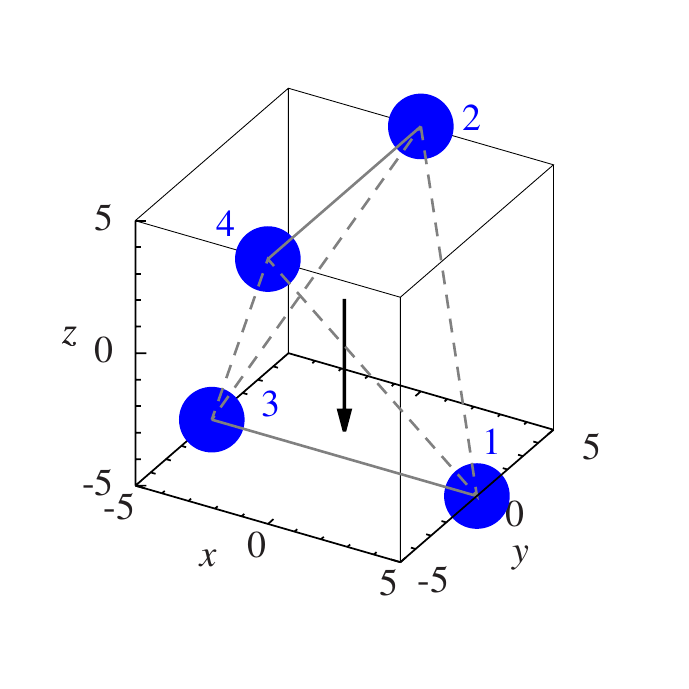} &
        \includegraphics[width=\linewidth]{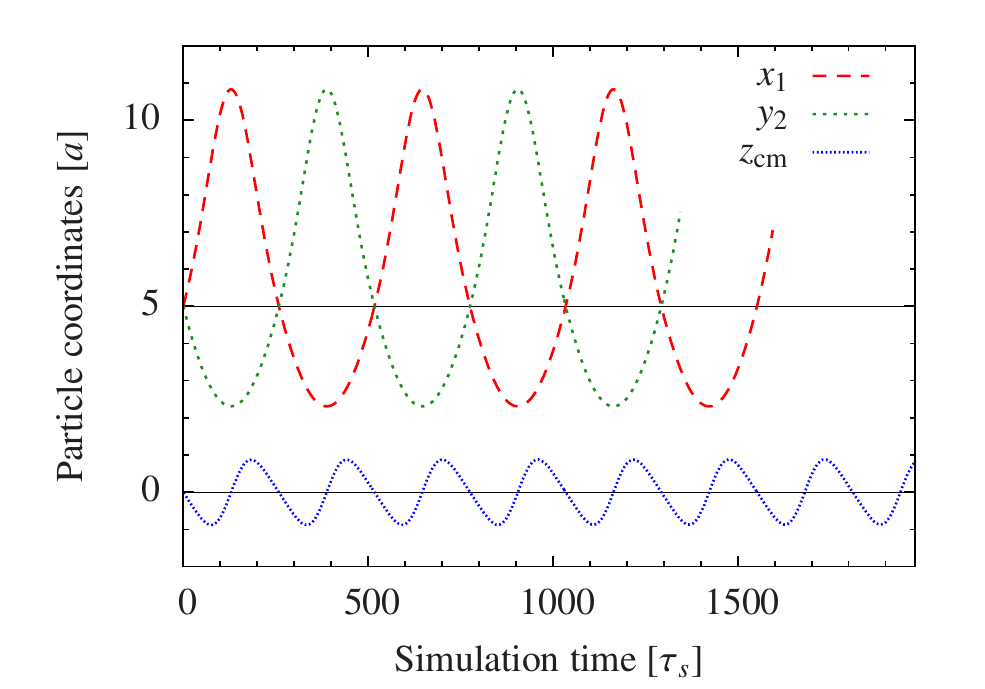}
    \end{tabular}
    \end{center}
    \caption{Periodic sedimentation of four equal spheres.  ~(a) Initial
    configuration of the particles on the edges of a stretched tetrahedron,
    edges of equal length are symbolised by equal line patterns.  Gravity acts
    downwards as indicated by the arrow.  ~(b) Periodic trajectories for two
    selected coordinates of particles 1 and 2 and the vertical component of the
    centre of mass, $z_\text{cm}$, after correcting for the average
    sedimentation motion.
    }
    \label{fig:periodic}
\end{figure}

In Stokesian dynamics simulations, there are no obvious conserved quantities
like energy or momentum, but one may exploit a symmetry to examine the numerical
stability of the simulation.
The Stokes equation~\eqref{eq:stokes} is invariant under time reversal
($\vec u \mapsto -\vec u$) if positions and forces are reversed as well,
$\vec r\mapsto -\vec r$, $\vec f\mapsto -\vec f$.
The symmetry of the Stokes flow is preserved for fore-and-aft-symmetric
particle configurations, i.e., for boundaries invariant under
$\vec r\mapsto -\vec r$.
These observations can be used to construct a cyclic motion in an external
gravitational field~\cite{Tory:1991}: four equal particles are initially
positioned at the vertices of a stretched tetrahedron with two opposing edges
perpendicular to the force, see \fig{periodic}a.
After a quarter period, all particles will be in a horizontal plane, and after
sedimenting for another quarter period, the horizontally mirrored initial
configuration is recovered.
After another half period, the initial (relative) configuration is exactly
restored.
For the initial positions, we placed two horizontally aligned and vertically
separated couples of particles at $\vec r_1 = (0,0,5a)$, $\vec r_2 = (0,5a,0)$,
$\vec r_3= -\vec r_1$, $\vec r_4= -\vec r_2$; each particle is dragged downwards
by a force $\vec F = (0, 0, -F)$.
The trajectories are determined by a set of merely three different coordinates,
e.g., $x_1(t)$, $z_1(t)$ and $z_2(t)$, see \fig{periodic}b, the remaining
components follow by symmetry: $x_3(t)=-x_1(t)$, $y_2(t) = -y_4(t)$ being equal to
$x_1(t)$ shifted in phase by half a period, $z_3(t)=z_1(t)$,  $z_4(t) = z_2(t)$, and $y_1
= y_3 = x_2 = x_4 \equiv 0$.
The centre-of-mass position, $z_{cm}(t)=[z_1(t) + z_2(t)]/2$, decreases monotonically, but
oscillates around the average sedimentation motion.

The cycle duration is about $T=517 \tau_s$ long, we have performed long
simulations over almost 1000 cycles (50 million steps with an integration
timestep of $0.01 \tau_s$) in order to monitor the numerical stability.
The maxima and minima of $x_1(t)$ were determined by parabolic fits yielding the
cycle duration and the amplitude of the motion, the cycle duration was averaged
over time windows of $10^4\tau_s$.
We compare implementations which use either single or double-single
floating-point precision \cite{Glassy_GPU:2011} for the accumulation of (a)
velocities due to hydrodynamic interactions,  \eq{rotneMobility}, and (b)
positions in the Euler integration of the equations of motion.
The contributions to the hydrodynamic interactions from different particles vary
strongly for close and distant particles due to the $1/r$ prefactor; hence,
large and small numbers of different sign are summed with great potential for
numerical issues due to limited floating-point precision.

\begin{figure}
  \begin{center}
    \includegraphics[width=\linewidth]{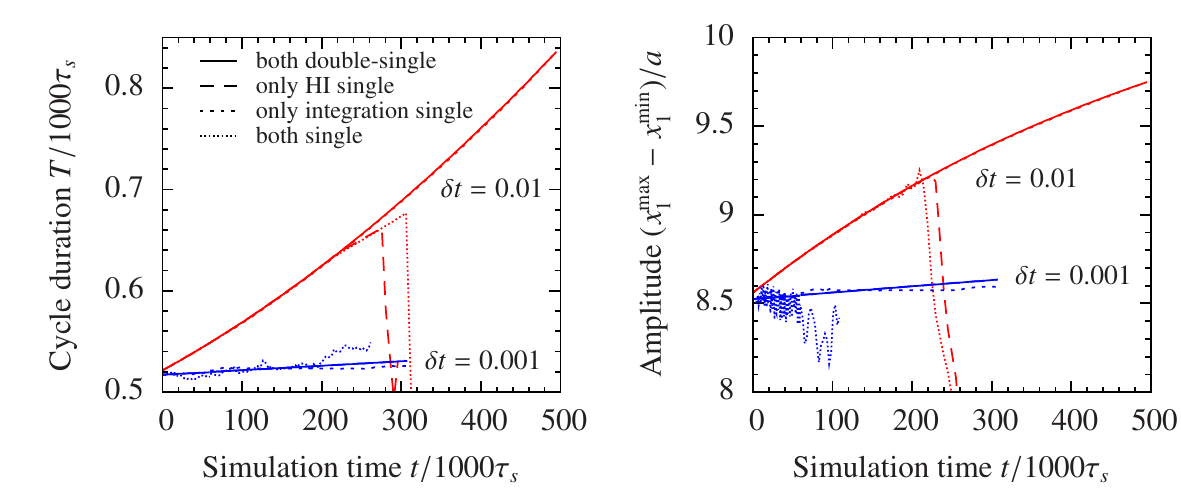}
  \end{center}
  \caption{Cycle duration and amplitude of the periodic motion over the course
  of a very long simulation run using different floating-point precision for the
  accumulation of hydrodynamic interactions (HI) and positions (integration of
  the equations of motion); the two datasets refer to different integration
  timesteps, $\delta t = 0.001 \tau_s$ and $\delta t = 0.01 \tau_s$ as
  indicated. Note that some curves collapse with the case ``both double-single
  precisison''.}
  \label{fig:precision}
\end{figure}

Over the course of the simulation, both the cycle duration and the amplitude
display a drift independently of the floating-point precision, see
\fig{precision}.
The motion, however, remains cyclic if double-single precision is used for the
accumulation of hydrodynamic interactions.
Otherwise, the periodic motion falls apart after some 400 cycles for the larger
timestep, regardless of the precision used for the integration.
Using the smaller timestep $\delta t = 0.001 \tau_s$ for the integration of
positions, the drift is significantly reduced and the cyclic motion remains
stable over the full simulation run of more than 500 cycles.
The small timestep in conjunction with single precision for both the
accumulation of hydrodynamic interactions and the integration leads to a wildly
fluctuating cycle duration and amplitude, probably because the low
floating-point precision fails to properly resolve the small increments.
Note that a smaller drift and an improved stability may be achieved by resorting
to higher-order or implicit integration schemes rather than decreasing the
timestep; a semi-implicit scheme has been suggested if lubrication forces are included~\cite{Nguyen:2002}.
The stability issues, however, are likely to persist if merely single floating-point
precision is used throughout.

\begin{figure}
  \begin{center}
    \includegraphics[width=\figwidth]{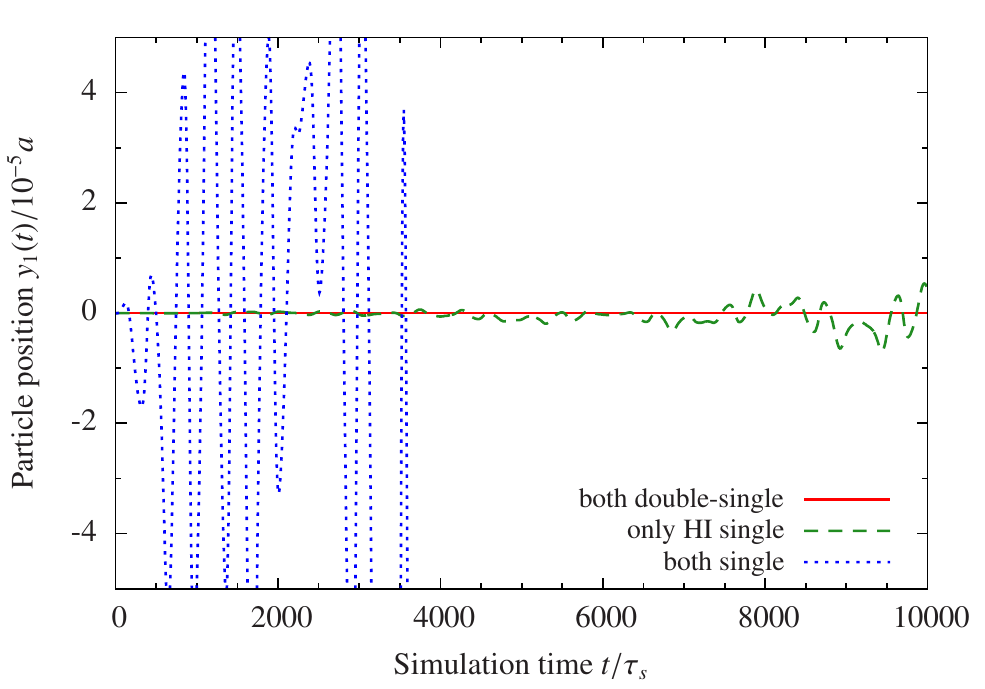}
  \end{center}
  \caption{Deviation of particle~1 from the $xz$-plane quantifying the numerical error.
  Accumulation with different floating-point precision is compared like in \fig{precision}.
  }
  \label{fig:precision2}
\end{figure}

Another sensitive criterion for numerical stability is that particles initially
positioned in the $xz$-plane (particles 1 and 3) should for symmetry reasons
remain in this plane during the simulation.
Thus, a non-zero $y$-coordinate quantifies the numerical error, the results are
shown in \fig{precision2} using $\delta t=0.001\tau_s$.
Note the much shorter simulation runs compared to \fig{precision}, the criterion
here is much more sensitive to detect numerical inaccuracies.
If double-single precision is used for the accumulation of hydrodynamic
interactions, we find $y_2(t)\equiv 0$, independently of the precision of the
integration.
Slowly growing deviations are observed if only the hydrodynamic interactions are
accumulated in single precision, and large errors arise quickly for positions
integrated in single precision as well.

In conclusion, the floating-point precision in the accumulation of hydrodynamic
interactions impacts the long-time stability of the described simulation for
only four particles, and it is likely to become more relevant for larger setups
with a large distribution of particle separations.
For the tests discussed, the precision of position integration is less crucial
as long as the hydrodynamic interactions are accumulated with increased
floating-point precision, but we consider it good practice to do so for the
integration as well.

\section{Conclusion}

We have described a GPU accelerated implementation of Stokesian dynamics
simulations, a scheme for dynamic simulations of a particulate suspension that
includes direct, instantaneous hydrodynamic interactions and avoids the explicit
simulation of the solvent.
Our treatment is restricted to the leading order of a multipole expansion, thus
neglecting rotational motion, but can in principle be extended to higher orders.
The implementation is available in \emph{HAL's MD package}~\cite{HALMD} from
version~0.2, an efficient simulation software specifically designed to run
entirely on the GPU, thereby avoiding the costly data transfer between host and
GPU memory; it further supports selected computations with increased
floating-point precision for an optimal balance between numerical stability and
runtime performance.
Since access to global memory on the GPU often limits the execution speed of
compute kernels, we have put particular emphasis on the memory access patterns.
Explicit data caching by means of the shared memory was compared to the hardware
cache available in the latest generation of Nvidia GPUs. Dedicated algorithms
using shared memory are in general more efficient for non-linear memory access.
For linear access, they show advantages for small data sizes, the straightforward
direct memory access, however, turned out to be an equally fast (or even
slightly faster) alternative for the computation of hydrodynamic interactions,
where a relatively large amount of data (32~bytes) has to be transferred for
each particle.

The computation of the long-range hydrodynamic interactions requires to consider
all pairs of particles in the simulation box, yielding an algorithmic complexity
of order $O\bigl(N^2\bigr)$ for $N$ particles.
This scaling is reflected in the runtime for large systems, and we observed a
speedup of about 120 compared to the serial reference implementation for a
conventional CPU.
For up to 2,000 particles however, the runtime scales linearly on the employed
GPU hardware since only a fraction of the available computing resources is
used.

Other simulation approaches for particulate suspensions treat the solvent
explicitly, which typically results in a linear algorithmic complexity.
For fixed particle number, the prefactor is proportional to the solvent volume,
or equally, the inverse of the colloid volume fraction.
\citet{Roehm:2012} describe molecular dynamics simulations of particles
suspended in a GPU-accelerated lattice-Boltzmann solvent.
For comparison, they report that one integration step takes about 0.9\,ms for
4,150 particles in a solvent of $24^3$ lattice nodes, which is to be contrasted
with 5.7\,ms obtained in this work.
Note that the execution time for Stokesian dynamics is merely a function of the
absolute particle number, while for the lattice-Boltzmann method it crucially
depends on the particle density and the resolution of the solvent lattice.
Hence, Stokesian dynamics will in general perform better for dilute suspensions.
Similarly, \citet{Pham:2009} compared both methods (including Brownian motion)
for a polymer chain in solution using conventional hardware and found Stokesian
dynamics to be an order of magnitude faster even for chain lengths of 1000
monomers, despite the asymptotically favourable scaling of the lattice-Boltzmann
method.
Apart from aspects of computational performance, a comparison of methods with
implicit and explicit solvent may be interesting from a theoretical point of
view for specific problems in physics.
It would permit assessing the physical quality of the various approximations
introduced by both methods.

The hydrodynamic interactions are related in nature to other long-range
interactions, e.g., gravity or electrostatic interactions, and it seems natural
to convey more elaborate algorithms from there to the hydrodynamic case,
an example being Ewald summation techniques~\cite{Sierou:2001}.
Another interesting approach are fast-multipole methods and tree codes
with $O(N\log(N))$ scaling, where remote
particle groups are replaced by their multipole moments;
Bédorf \emph{et al.}~\cite{Bedorf:2012, Bedorf:2012a} describe an implementation
of a Barnes--Hut tree code for gravity fully executing on the GPU, which is able
to process 2.8~million particles per second.

Finally, one would like to amend the simulations by Brownian
motion~\cite{Ermak:1978, Banchio:2003}, which involves an additional noise term.
In order to satisfy the detailed balance condition, a correct computation of the
noise strength requires a Cholesky decomposition of the many-body mobility
matrix $\tensor \textmu$, which is computationally expensive and entails a
substantial memory footprint.
Nevertheless, such a task seems to be well suited for a future GPU
implementation.

\begin{acknowledgement}
  We thank J.\ Dunkel, T.\ Franosch, and A.\ Louis for discussions on Stokesian
  dynamics simulations and hydrodynamic interactions and P.\ Colberg for
  correspondence on GPU-related questions.
\end{acknowledgement}

\appendix

\section{~~Data caching for memory access patterns of complexity \boldmath $O\bigl(N^2\bigr)$}
\label{sec:compAlgo}

\begin{figure}
  \centering
  \includegraphics[width=\figwidth]{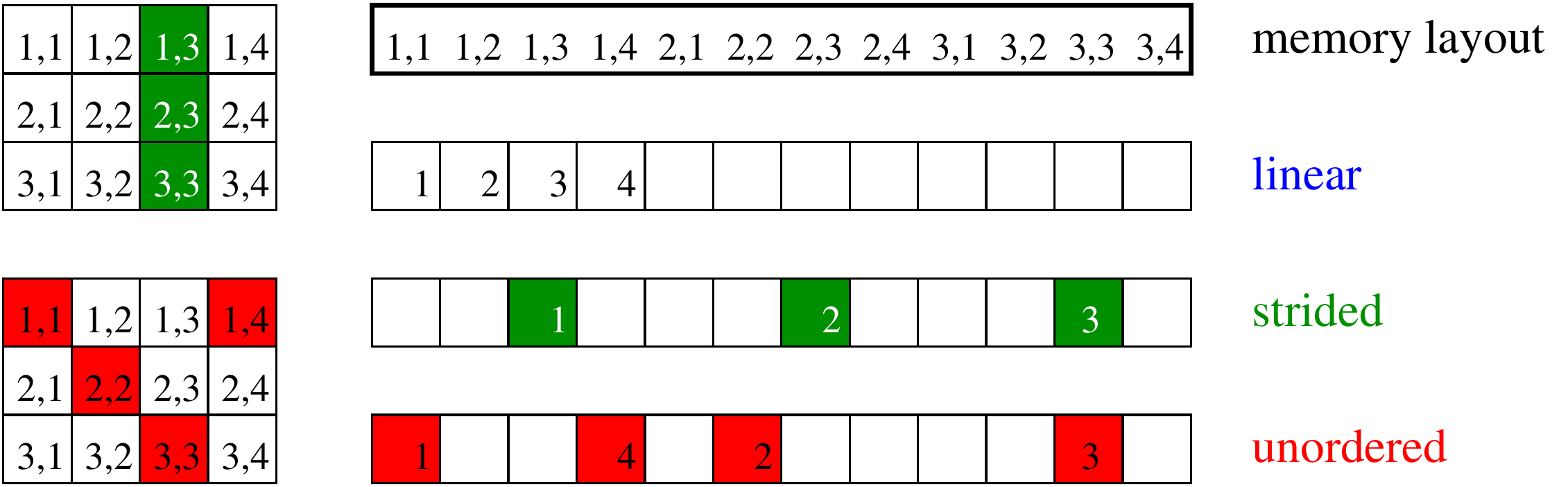}
  \caption{Illustration of the different access patterns.  Data are assumed
  to represent a $3 \times 4$ matrix (left) stored in row-major order (white bar on top).
  ``Linear'' data access corresponds to reading a \emph{row} of the matrix,
  ``strided'' access reads a column, and ``unordered'' access reads along a
  diagonal depending on the offset. Single numbers denote the sequence of data
  access, and comma-separated pairs indices of matrix elements.}
  \label{fig:accesspatterns}
\end{figure}

In this appendix, we study the efficiency of data caching using shared memory or
hardware caches for different memory access patterns of complexity
$O\bigl(N^2\bigr)$, which occur in the computation of all-pairs interactions or
in basic linear algebra problems (BLAS).
The study was motivated by the somewhat unexpected observation that the two
implementations of global memory access in Algorithms~\ref{algo:naiveKernel} and
\ref{algo:smalltilingKernel} yielded similar execution times, even on older GPU
hardware without memory cache.
Three different memory access patterns, depicted in \fig{accesspatterns}, are
addressed: linear, linear with a fixed stride, and unordered scattered.

\begin{algorithm}[b]
  \caption{Minimalist na{\"i}ve kernel for benchmarking memory access.}
  \label{algo:miniNaive}
  \begin{verbatim}
template <typename T>
__global__ void naive(T* g_out, T const* g_in, unsigned N)
{
    unsigned const j = GTID;                     // global thread ID
    T sum = g_in[j];                             // read from global memory
    for (unsigned k = 0; k < N; ++k)
        sum += g_in[map(k)];                     // read from global memory
                                                 // and a trivial computation
    g_out[j] = sum;                              // write to global memory
} \end{verbatim}
\end{algorithm}

\begin{algorithm}[b]
  \caption{Minimalist small-tiling kernel.  The ``tiling''
  algorithm is obtained by substituting \texttt{warpSize} with the
  size of the execution blocks; the condition \texttt{n < warpSize} in line~10
  then always evaluates to true and can be dropped.}
  \label{algo:miniSmalltiling}
  \begin{verbatim}
template <typename T>
__global__ void smalltiling(T* g_out, T const* g_in, unsigned N)
{
    unsigned const j = GTID;                     // global thread ID
    unsigned const n = TID;                      // thread ID within block
    extern __shared__ T s_data[];                // declare shared memory
    T sum = g_in[j];                             // read from global memory

    for (unsigned m = 0; m < N; m += warpSize) {
        if (n < warpSize)
            s_data[n] = g_in[map(m+n)];          // copy from global to
        __syncthreads();                         // shared memory

        for (unsigned k = 0; k < warpSize; ++k)
            sum += s_data[k];                    // a trivial computation
        __syncthreads();
    }
    g_out[j] = sum;                              // write to global memory
} \end{verbatim}
\end{algorithm}

We reduced the algorithms of Section~\ref{sec:impl} as much as possible to
concentrate on memory access.
A minimal computation though was included in order to prevent the compiler from
optimising out superfluous read statements.
The three minimalist CUDA kernels are given in Algorithms~\ref{algo:miniNaive}
and \ref{algo:miniSmalltiling}, also see \fig{implementations}.
The kernels are run with $N$ threads, each thread reads and accumulates an array
of $N$ values and stores the result.
The \textsf{naive} kernel resembles the straightforward implementation,
\textsf{smalltiling} and \textsf{tiling} divide the array in tiles and use
shared memory for explicit data caching.
The memory access patterns were realised by the mappings $k \mapsto k$ for
linear access, $k \mapsto k p \!\mod N$ for strided access, and $k \mapsto (k p
+ j M) \!\mod N$ for unordered access; $k$ is the index of the memory
access loop, $p = 131$ a prime number larger than 128, $M=32$ the warp size, and
$j$ the global thread ID.
For linear access, the threads of a warp request data from adjacent locations
permitting coalescable memory transfer.
For the strided access pattern, two ``consecutive'' threads access data
separated by $p \times s$ bytes ($s$ being the size of the basic array data
type), thus two array elements can not be fetched by a single transaction.
For the unordered access, each thread traverses the data array with a unique
offset preventing different threads to benefit from processing the same data.
The three patterns are reminiscent of traversing a $p\times p$ matrix stored in
row-major order in different directions: reading a row (linear), a column
(strided), or a diagonal (unordered), see \fig{accesspatterns}.

\begin{figure}
  \centering
  \includegraphics[width=.49\linewidth]{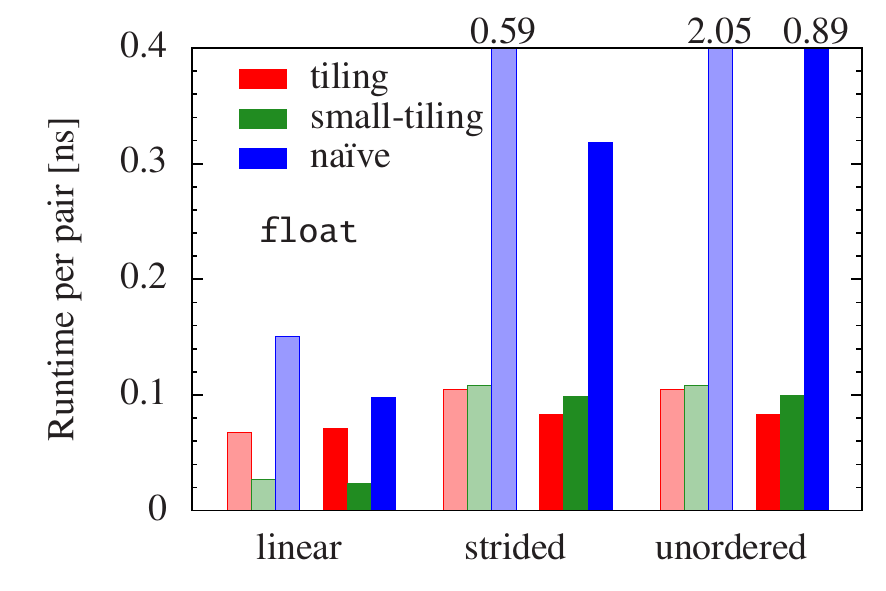} \hfill
  \includegraphics[width=.49\linewidth]{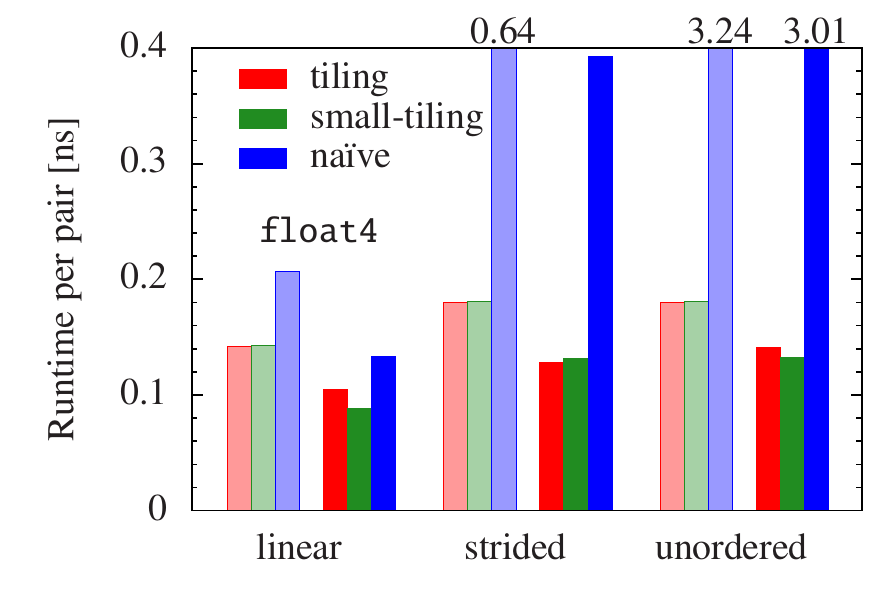}
  \caption{Runtime performance of different $O\bigl(N^2\bigr)$-memory access
  patterns using various GPU implementations, see
  Algorithms~\ref{algo:miniNaive} and \ref{algo:miniSmalltiling}.  The figures
  show kernel execution times divided by the number of pairs, $N^2$, in the
  scaling limit of a large number of array elements, $N$.  The two panels
  compare arrays of data type \textsf{float} (left) and \textsf{float4} (right).
  Light (left groups) and solid (right groups) fillings distinguish results for
  GPUs without and with hardware cache, respectively; see text for details.
  Some runtimes of the na\"ive implementation exceed the axis range and are
  printed at the top.
  Note that the total amount of transferred memory is larger by a
  factor of 4 for \textsf{float4} values.
  }
  \label{fig:scaling}
\end{figure}

Runtime measurements were conducted on two systems with GPU hardware with and
without on-chip memory cache.  The first system hosted an Nvidia Tesla S1070
server (T10 chip, CUDA compute capability 1.3), and the second system contained
four Tesla C2050 (``Fermi'') cards (compute capability 2.0).  The latter GPU is,
among other improvements, equipped with a hardware cache of 48\,kB.  For both
systems, we used the same kernel builds created by the CUDA SDK version 3.2 with
the compilation option \textsf{--arch sm\_12}.  Kernels were run with 128
threads per block on a single GPU; kernel execution times were averaged over 100
repetitions and do not include data transfer from host to GPU memory.  The array
size was varied between $N=1{,}000$ and 70,000 checking that the runtime divided
by $N^2$ has converged, and the obtained runtimes are presented in
\fig{scaling}.

The runtime of the na\"ive implementation depends sensitively on the access
pattern, which nicely illustrates the effect of the hardware cache. In the
absence of a cache and for arrays of data type \textsf{float}, the
kernel with unordered access is slower by a factor of 13 compared to the linear
pattern; this reduces to 9 on the C2050 card. The factor, however, increases
from 15 (S1070) to 23 (C2050) if the basic data type is \textsf{float4}
(16~bytes per array element). Keeping the access pattern and the data type
fixed, the largest observed runtime improvement was by a factor of 2.3 for
unordered access and \textsf{float} data, which we mainly attribute to the
presence of memory cache and the higher number of multiprocessors on the C2050
GPU. The overall observation of a higher performance gain using the latter
hardware for \textsf{float} data is likely due to the fact that the
cache can store simply more \textsf{float} than \textsf{float4} data items,
reducing the number of cache misses upon later access of the same data by
another thread.

The two tiling algorithms are between 1.3 (linear access) to 20 times (unordered
access) faster than their respective na\"ive implementations.  Both algorithms
exhibit similar execution times, they are even comparable for the different
access patterns, solely the linear access is somewhat faster.  This shows that
manual caching via shared memory can efficiently hide even complex memory
access.  Comparing the two GPU generations, the C2050 is found to be slightly
more efficient for \textsf{float4} data, but to perform similarly as one of the
S1070 GPUs for \textsf{float}.  A notable exception is linear access to
\textsf{float} data, where the ``small-tiling'' implementation is about 2--3
times faster than the original tiling algorithm~\cite{CUDA_Guide, Nyland:2007}.
The advantages of the ``small-tiling'' version are expected to become relevant
only for larger array elements (e.g., 2\,\texttimes\,\textsf{float4}), when the
considerations on shared memory and multiprocessor occupancy given in
Section~\ref{sec:impl} apply.

Summarising, the built-in memory cache of the Tesla C2050 hardware yields
significant improvements to global memory access with the straightforward,
``na\"ive'' implementation of our test kernels.  In particular, this
implementation performs well for linear access of \textsf{float4} data in the
presence of a hardware cache.  It can, however, not compete with an explicit
caching algorithm via shared memory, which can be designed to reflect the actual
data processing scheme for optimal cache efficiency and coalescable memory
transfer.  The ``small-tiling'' algorithm achieves in all test cases a high
memory transfer bandwidth and seems to be rather robust with respect to access
patterns and sizes of the array elements.
It may be worth to investigate whether it can speed up matrix-vector
multiplications, where the data of the vector are linearly read by all threads.


\begin{thebibliography}{39}
\expandafter\ifx\csname natexlab\endcsname\relax\def\natexlab#1{#1}\fi
\expandafter\ifx\csname bibnamefont\endcsname\relax
  \def\bibnamefont#1{#1}\fi
\expandafter\ifx\csname bibfnamefont\endcsname\relax
  \def\bibfnamefont#1{#1}\fi
\expandafter\ifx\csname citenamefont\endcsname\relax
  \def\citenamefont#1{#1}\fi
\providecommand{\bibinfo}[2]{#2}

\bibitem[{\citenamefont{Franosch et~al.}(2011)\citenamefont{Franosch, Grimm,
  Belushkin, Mor, Foffi, Forro, and Jeney}}]{Franosch:2011}
\bibinfo{author}{\bibfnamefont{T.}~\bibnamefont{Franosch}},
  \bibinfo{author}{\bibfnamefont{M.}~\bibnamefont{Grimm}},
  \bibinfo{author}{\bibfnamefont{M.}~\bibnamefont{Belushkin}},
  \bibinfo{author}{\bibfnamefont{F.~M.} \bibnamefont{Mor}},
  \bibinfo{author}{\bibfnamefont{G.}~\bibnamefont{Foffi}},
  \bibinfo{author}{\bibfnamefont{L.}~\bibnamefont{Forro}}, \bibnamefont{and}
  \bibinfo{author}{\bibfnamefont{S.}~\bibnamefont{Jeney}},
  \bibinfo{journal}{Nature} \textbf{\bibinfo{volume}{478}}, \bibinfo{pages}{85}
  (\bibinfo{year}{2011}).

\bibitem[{\citenamefont{Lutz et~al.}(2006)\citenamefont{Lutz, Reichert, Stark,
  and Bechinger}}]{Lutz:2006}
\bibinfo{author}{\bibfnamefont{C.}~\bibnamefont{Lutz}},
  \bibinfo{author}{\bibfnamefont{M.}~\bibnamefont{Reichert}},
  \bibinfo{author}{\bibfnamefont{H.}~\bibnamefont{Stark}}, \bibnamefont{and}
  \bibinfo{author}{\bibfnamefont{C.}~\bibnamefont{Bechinger}},
  \bibinfo{journal}{Europhys. Lett.} \textbf{\bibinfo{volume}{74}},
  \bibinfo{pages}{719} (\bibinfo{year}{2006}).

\bibitem[{\citenamefont{Wysocki et~al.}(2009)\citenamefont{Wysocki, Royall,
  Winkler, Gompper, Tanaka, van Blaaderen, and L\"owen}}]{Wysocki:2009}
\bibinfo{author}{\bibfnamefont{A.}~\bibnamefont{Wysocki}},
  \bibinfo{author}{\bibfnamefont{C.~P.} \bibnamefont{Royall}},
  \bibinfo{author}{\bibfnamefont{R.~G.} \bibnamefont{Winkler}},
  \bibinfo{author}{\bibfnamefont{G.}~\bibnamefont{Gompper}},
  \bibinfo{author}{\bibfnamefont{H.}~\bibnamefont{Tanaka}},
  \bibinfo{author}{\bibfnamefont{A.}~\bibnamefont{van Blaaderen}},
  \bibnamefont{and} \bibinfo{author}{\bibfnamefont{H.}~\bibnamefont{L\"owen}},
  \bibinfo{journal}{Soft Matter} \textbf{\bibinfo{volume}{5}},
  \bibinfo{pages}{1340} (\bibinfo{year}{2009}).

\bibitem[{\citenamefont{Milinkovic et~al.}(2011)\citenamefont{Milinkovic,
  Padding, and Dijkstra}}]{Milinkovic:2011}
\bibinfo{author}{\bibfnamefont{K.}~\bibnamefont{Milinkovic}},
  \bibinfo{author}{\bibfnamefont{J.~T.} \bibnamefont{Padding}},
  \bibnamefont{and} \bibinfo{author}{\bibfnamefont{M.}~\bibnamefont{Dijkstra}},
  \bibinfo{journal}{Soft Matter} \textbf{\bibinfo{volume}{7}},
  \bibinfo{pages}{11177} (\bibinfo{year}{2011}).

\bibitem[{\citenamefont{Rinn et~al.}(1999)\citenamefont{Rinn, Zahn, Maass, and
  Maret}}]{Rinn:1999}
\bibinfo{author}{\bibfnamefont{B.}~\bibnamefont{Rinn}},
  \bibinfo{author}{\bibfnamefont{K.}~\bibnamefont{Zahn}},
  \bibinfo{author}{\bibfnamefont{P.}~\bibnamefont{Maass}}, \bibnamefont{and}
  \bibinfo{author}{\bibfnamefont{G.}~\bibnamefont{Maret}},
  \bibinfo{journal}{Europhys. Lett.} \textbf{\bibinfo{volume}{46}},
  \bibinfo{pages}{537} (\bibinfo{year}{1999}).

\bibitem[{\citenamefont{Bleibel et~al.}(2011)\citenamefont{Bleibel, Dietrich,
  Dom\'inguez, and Oettel}}]{Bleibel:2011}
\bibinfo{author}{\bibfnamefont{J.}~\bibnamefont{Bleibel}},
  \bibinfo{author}{\bibfnamefont{S.}~\bibnamefont{Dietrich}},
  \bibinfo{author}{\bibfnamefont{A.}~\bibnamefont{Dom\'inguez}},
  \bibnamefont{and} \bibinfo{author}{\bibfnamefont{M.}~\bibnamefont{Oettel}},
  \bibinfo{journal}{Phys. Rev. Lett.} \textbf{\bibinfo{volume}{107}},
  \bibinfo{pages}{128302} (\bibinfo{year}{2011}).

\bibitem[{\citenamefont{Dünweg and Ladd}(2009)}]{Duenweg:2009}
\bibinfo{author}{\bibfnamefont{B.}~\bibnamefont{Dünweg}} \bibnamefont{and}
  \bibinfo{author}{\bibfnamefont{A.}~\bibnamefont{Ladd}}, in
  \emph{\bibinfo{booktitle}{Advanced Computer Simulation Approaches for Soft
  Matter Sciences {III}}}, edited by
  \bibinfo{editor}{\bibfnamefont{C.}~\bibnamefont{Holm}} \bibnamefont{and}
  \bibinfo{editor}{\bibfnamefont{K.}~\bibnamefont{Kremer}}
  (\bibinfo{publisher}{Springer Berlin/Heidelberg}, \bibinfo{year}{2009}), vol.
  \bibinfo{volume}{221} of \emph{\bibinfo{series}{Advances in Polymer
  Science}}, pp. \bibinfo{pages}{89--166}, ISBN
  \bibinfo{isbn}{978-3-540-87705-9}.

\bibitem[{\citenamefont{Gompper et~al.}(2009)\citenamefont{Gompper, Ihle,
  Kroll, and Winkler}}]{Gompper:2009}
\bibinfo{author}{\bibfnamefont{G.}~\bibnamefont{Gompper}},
  \bibinfo{author}{\bibfnamefont{T.}~\bibnamefont{Ihle}},
  \bibinfo{author}{\bibfnamefont{D.}~\bibnamefont{Kroll}}, \bibnamefont{and}
  \bibinfo{author}{\bibfnamefont{R.}~\bibnamefont{Winkler}}, in
  \emph{\bibinfo{booktitle}{Advanced Computer Simulation Approaches for Soft
  Matter Sciences {III}}}, edited by
  \bibinfo{editor}{\bibfnamefont{C.}~\bibnamefont{Holm}} \bibnamefont{and}
  \bibinfo{editor}{\bibfnamefont{K.}~\bibnamefont{Kremer}}
  (\bibinfo{publisher}{Springer Berlin/Heidelberg}, \bibinfo{year}{2009}), vol.
  \bibinfo{volume}{221} of \emph{\bibinfo{series}{Advances in Polymer
  Science}}, pp. \bibinfo{pages}{1--87}, ISBN
  \bibinfo{isbn}{978-3-540-87705-9}.

\bibitem[{\citenamefont{Padding and Louis}(2006)}]{Padding:2006}
\bibinfo{author}{\bibfnamefont{J.~T.} \bibnamefont{Padding}} \bibnamefont{and}
  \bibinfo{author}{\bibfnamefont{A.~A.} \bibnamefont{Louis}},
  \bibinfo{journal}{Phys. Rev. E} \textbf{\bibinfo{volume}{74}},
  \bibinfo{pages}{031402} (\bibinfo{year}{2006}).

\bibitem[{\citenamefont{Ermak and McCammon}(1978)}]{Ermak:1978}
\bibinfo{author}{\bibfnamefont{D.~L.} \bibnamefont{Ermak}} \bibnamefont{and}
  \bibinfo{author}{\bibfnamefont{J.~A.} \bibnamefont{McCammon}},
  \bibinfo{journal}{J. Chem. Phys.} \textbf{\bibinfo{volume}{69}},
  \bibinfo{pages}{1352} (\bibinfo{year}{1978}).

\bibitem[{\citenamefont{Durlofsky et~al.}(1987)\citenamefont{Durlofsky, Brady,
  and Bossis}}]{Durlofsky:1987}
\bibinfo{author}{\bibfnamefont{L.}~\bibnamefont{Durlofsky}},
  \bibinfo{author}{\bibfnamefont{J.~F.} \bibnamefont{Brady}}, \bibnamefont{and}
  \bibinfo{author}{\bibfnamefont{G.}~\bibnamefont{Bossis}},
  \bibinfo{journal}{J. Fluid Mech.} \textbf{\bibinfo{volume}{180}},
  \bibinfo{pages}{21} (\bibinfo{year}{1987}).

\bibitem[{\citenamefont{Banchio and Brady}(2003)}]{Banchio:2003}
\bibinfo{author}{\bibfnamefont{A.~J.} \bibnamefont{Banchio}} \bibnamefont{and}
  \bibinfo{author}{\bibfnamefont{J.~F.} \bibnamefont{Brady}},
  \bibinfo{journal}{J. Chem. Phys.} \textbf{\bibinfo{volume}{118}},
  \bibinfo{pages}{10323} (\bibinfo{year}{2003}).

\bibitem[{\citenamefont{Ladd}(1993)}]{Ladd:1993}
\bibinfo{author}{\bibfnamefont{A.~J.~C.} \bibnamefont{Ladd}},
  \bibinfo{journal}{Phys. Fluids A} \textbf{\bibinfo{volume}{5}},
  \bibinfo{pages}{299} (\bibinfo{year}{1993}).

\bibitem[{\citenamefont{Padding and Louis}(2004)}]{Padding:2004}
\bibinfo{author}{\bibfnamefont{J.~T.} \bibnamefont{Padding}} \bibnamefont{and}
  \bibinfo{author}{\bibfnamefont{A.~A.} \bibnamefont{Louis}},
  \bibinfo{journal}{Phys. Rev. Lett.} \textbf{\bibinfo{volume}{93}},
  \bibinfo{pages}{220601} (\bibinfo{year}{2004}).

\bibitem[{\citenamefont{Putz et~al.}(2010)\citenamefont{Putz, Dunkel, and
  Yeomans}}]{Putz:2010}
\bibinfo{author}{\bibfnamefont{V.~B.} \bibnamefont{Putz}},
  \bibinfo{author}{\bibfnamefont{J.}~\bibnamefont{Dunkel}}, \bibnamefont{and}
  \bibinfo{author}{\bibfnamefont{J.~M.} \bibnamefont{Yeomans}},
  \bibinfo{journal}{Chem. Phys.} \textbf{\bibinfo{volume}{375}},
  \bibinfo{pages}{557 } (\bibinfo{year}{2010}).

\bibitem[{\citenamefont{Baker and Hirst}(2011)}]{Baker:2011}
\bibinfo{author}{\bibfnamefont{J.~A.} \bibnamefont{Baker}} \bibnamefont{and}
  \bibinfo{author}{\bibfnamefont{J.~D.} \bibnamefont{Hirst}},
  \bibinfo{journal}{Mol. Inform.} \textbf{\bibinfo{volume}{30}},
  \bibinfo{pages}{498} (\bibinfo{year}{2011}).

\bibitem[{\citenamefont{Anderson et~al.}(2008)\citenamefont{Anderson, Lorenz,
  and Travesset}}]{Anderson:2008}
\bibinfo{author}{\bibfnamefont{J.~A.} \bibnamefont{Anderson}},
  \bibinfo{author}{\bibfnamefont{C.~D.} \bibnamefont{Lorenz}},
  \bibnamefont{and}
  \bibinfo{author}{\bibfnamefont{A.}~\bibnamefont{Travesset}},
  \bibinfo{journal}{J. Comp. Phys.} \textbf{\bibinfo{volume}{227}},
  \bibinfo{pages}{5342} (\bibinfo{year}{2008}).

\bibitem[{\citenamefont{van Meel et~al.}(2008)\citenamefont{van Meel, Arnold,
  Frenkel, Portegies~Zwart, and Belleman}}]{Meel:2008}
\bibinfo{author}{\bibfnamefont{J.~A.} \bibnamefont{van Meel}},
  \bibinfo{author}{\bibfnamefont{A.}~\bibnamefont{Arnold}},
  \bibinfo{author}{\bibfnamefont{D.}~\bibnamefont{Frenkel}},
  \bibinfo{author}{\bibfnamefont{S.}~\bibnamefont{Portegies~Zwart}},
  \bibnamefont{and} \bibinfo{author}{\bibfnamefont{R.~G.}
  \bibnamefont{Belleman}}, \bibinfo{journal}{Mol. Simul.}
  \textbf{\bibinfo{volume}{34}}, \bibinfo{pages}{259} (\bibinfo{year}{2008}).

\bibitem[{\citenamefont{Colberg and H{\"o}f{}ling}(2011)}]{Glassy_GPU:2011}
\bibinfo{author}{\bibfnamefont{P.~H.} \bibnamefont{Colberg}} \bibnamefont{and}
  \bibinfo{author}{\bibfnamefont{F.}~\bibnamefont{H{\"o}f{}ling}},
  \bibinfo{journal}{Comput. Phys. Commun.} \textbf{\bibinfo{volume}{182}},
  \bibinfo{pages}{1120} (\bibinfo{year}{2011}).

\bibitem[{\citenamefont{Ruymgaart et~al.}(2011)\citenamefont{Ruymgaart,
  Cardenas, and Elber}}]{Ruymgaart:2011}
\bibinfo{author}{\bibfnamefont{A.~P.} \bibnamefont{Ruymgaart}},
  \bibinfo{author}{\bibfnamefont{A.~E.} \bibnamefont{Cardenas}},
  \bibnamefont{and} \bibinfo{author}{\bibfnamefont{R.}~\bibnamefont{Elber}},
  \bibinfo{journal}{J. Chem. Theory Comput.} \textbf{\bibinfo{volume}{7}},
  \bibinfo{pages}{3072} (\bibinfo{year}{2011}).

\bibitem[{\citenamefont{R\"ohm and Arnold}(2012)}]{Roehm:2012}
\bibinfo{author}{\bibfnamefont{D.}~\bibnamefont{R\"ohm}} \bibnamefont{and}
  \bibinfo{author}{\bibfnamefont{A.}~\bibnamefont{Arnold}},
  \bibinfo{journal}{Eur. Phys. J. Special Topics}  (\bibinfo{year}{2012}),
  \bibinfo{note}{in this issue}.

\bibitem[{\citenamefont{Batchelor}(2000)}]{Batchelor:FluidDynamics}
\bibinfo{author}{\bibfnamefont{G.}~\bibnamefont{Batchelor}},
  \emph{\bibinfo{title}{An introduction to fluid dynamics}}, Cambridge
  mathematical library (\bibinfo{publisher}{Cambridge University Press},
  \bibinfo{year}{2000}), ISBN \bibinfo{isbn}{9780521663960}.

\bibitem[{\citenamefont{Dhont}(1996)}]{Dhont:ColloidDynamics}
\bibinfo{author}{\bibfnamefont{J.~K.~G.} \bibnamefont{Dhont}},
  \emph{\bibinfo{title}{An introduction to dynamics of colloids}}
  (\bibinfo{publisher}{Elsevier}, \bibinfo{year}{1996}), \bibinfo{edition}{2nd}
  ed.

\bibitem[{\citenamefont{Eckhardt and Buehrle}(2008)}]{Eckhardt:2008}
\bibinfo{author}{\bibfnamefont{B.}~\bibnamefont{Eckhardt}} \bibnamefont{and}
  \bibinfo{author}{\bibfnamefont{J.}~\bibnamefont{Buehrle}},
  \bibinfo{journal}{Eur. Phys. J. Special Topics}
  \textbf{\bibinfo{volume}{157}}, \bibinfo{pages}{135} (\bibinfo{year}{2008}).

\bibitem[{\citenamefont{Oseen}(1927)}]{Oseen:Hydrodynamik}
\bibinfo{author}{\bibfnamefont{C.~W.} \bibnamefont{Oseen}},
  \emph{\bibinfo{title}{Neuere Methoden und Ergebnisse in der Hydrodynamik}}
  (\bibinfo{publisher}{Akad. Verlagsges.}, \bibinfo{address}{Leipzig},
  \bibinfo{year}{1927}).

\bibitem[{\citenamefont{Mazur and van Saarloos}(1982)}]{Mazur:1982}
\bibinfo{author}{\bibfnamefont{P.}~\bibnamefont{Mazur}} \bibnamefont{and}
  \bibinfo{author}{\bibfnamefont{W.}~\bibnamefont{van Saarloos}},
  \bibinfo{journal}{Physica A} \textbf{\bibinfo{volume}{115}},
  \bibinfo{pages}{21 } (\bibinfo{year}{1982}).

\bibitem[{\citenamefont{Rotne and Prager}(1969)}]{Rotne:1969}
\bibinfo{author}{\bibfnamefont{J.}~\bibnamefont{Rotne}} \bibnamefont{and}
  \bibinfo{author}{\bibfnamefont{S.}~\bibnamefont{Prager}},
  \bibinfo{journal}{J. Chem. Phys.} \textbf{\bibinfo{volume}{50}},
  \bibinfo{pages}{4831} (\bibinfo{year}{1969}).

\bibitem[{\citenamefont{Cox}(1974)}]{Cox:1974}
\bibinfo{author}{\bibfnamefont{R.}~\bibnamefont{Cox}}, \bibinfo{journal}{Int.
  J. Multiphase Flow} \textbf{\bibinfo{volume}{1}}, \bibinfo{pages}{343 }
  (\bibinfo{year}{1974}).

\bibitem[{\citenamefont{Jeffrey and Onishi}(1984)}]{Jeffrey:1984}
\bibinfo{author}{\bibfnamefont{D.~J.} \bibnamefont{Jeffrey}} \bibnamefont{and}
  \bibinfo{author}{\bibfnamefont{Y.}~\bibnamefont{Onishi}},
  \bibinfo{journal}{J. Fluid Mech.} \textbf{\bibinfo{volume}{139}},
  \bibinfo{pages}{261} (\bibinfo{year}{1984}).

\bibitem[{\citenamefont{Ladd}(1990)}]{Ladd:1990}
\bibinfo{author}{\bibfnamefont{A.~J.~C.} \bibnamefont{Ladd}},
  \bibinfo{journal}{J. Chem. Phys.} \textbf{\bibinfo{volume}{93}},
  \bibinfo{pages}{3484} (\bibinfo{year}{1990}).

\bibitem[{CUD(2010)}]{CUDA_Guide}
\emph{\bibinfo{title}{{NVIDIA} {CUDA} {C} {P}rogramming {G}uide}}
  (\bibinfo{year}{2010}), \bibinfo{note}{version 3.2}.

\bibitem[{\citenamefont{Nyland et~al.}(2007)\citenamefont{Nyland, Harris, and
  Prins}}]{Nyland:2007}
\bibinfo{author}{\bibfnamefont{L.}~\bibnamefont{Nyland}},
  \bibinfo{author}{\bibfnamefont{M.}~\bibnamefont{Harris}}, \bibnamefont{and}
  \bibinfo{author}{\bibfnamefont{J.}~\bibnamefont{Prins}}, in
  \emph{\bibinfo{booktitle}{{GPU} Gems 3}}, edited by
  \bibinfo{editor}{\bibfnamefont{H.}~\bibnamefont{Nguyen}}
  (\bibinfo{publisher}{Addison Wesley}, \bibinfo{year}{2007}),
  chap.~\bibinfo{chapter}{31}.

\bibitem[{\citenamefont{Colberg and H{\"o}f{}ling}(2008--2012)}]{HALMD}
\bibinfo{author}{\bibfnamefont{P.~H.} \bibnamefont{Colberg}} \bibnamefont{and}
  \bibinfo{author}{\bibfnamefont{F.}~\bibnamefont{H{\"o}f{}ling}},
  \emph{\bibinfo{title}{{H}ighly {A}ccelerated {L}arge-scale {M}olecular
  {D}ynamics package}} (\bibinfo{year}{2008--2012}),
  \bibinfo{note}{\url{http://halmd.org}}.

\bibitem[{\citenamefont{Tory et~al.}(1991)\citenamefont{Tory, Kamel, and
  Tory}}]{Tory:1991}
\bibinfo{author}{\bibfnamefont{E.}~\bibnamefont{Tory}},
  \bibinfo{author}{\bibfnamefont{M.}~\bibnamefont{Kamel}}, \bibnamefont{and}
  \bibinfo{author}{\bibfnamefont{C.}~\bibnamefont{Tory}},
  \bibinfo{journal}{Powder Technol.} \textbf{\bibinfo{volume}{67}},
  \bibinfo{pages}{71 } (\bibinfo{year}{1991}).

\bibitem[{\citenamefont{Nguyen and Ladd}(2002)}]{Nguyen:2002}
\bibinfo{author}{\bibfnamefont{N.-Q.} \bibnamefont{Nguyen}} \bibnamefont{and}
  \bibinfo{author}{\bibfnamefont{A.~J.~C.} \bibnamefont{Ladd}},
  \bibinfo{journal}{Phys. Rev. E} \textbf{\bibinfo{volume}{66}},
  \bibinfo{pages}{046708} (\bibinfo{year}{2002}).

\bibitem[{\citenamefont{Pham et~al.}(2009)\citenamefont{Pham, Schiller,
  Prakash, and D\"unweg}}]{Pham:2009}
\bibinfo{author}{\bibfnamefont{T.~T.} \bibnamefont{Pham}},
  \bibinfo{author}{\bibfnamefont{U.~D.} \bibnamefont{Schiller}},
  \bibinfo{author}{\bibfnamefont{J.~R.} \bibnamefont{Prakash}},
  \bibnamefont{and} \bibinfo{author}{\bibfnamefont{B.}~\bibnamefont{D\"unweg}},
  \bibinfo{journal}{J. Chem. Phys.} \textbf{\bibinfo{volume}{131}},
  \bibinfo{pages}{164114} (\bibinfo{year}{2009}).

\bibitem[{\citenamefont{Sierou and Brady}(2001)}]{Sierou:2001}
\bibinfo{author}{\bibfnamefont{A.}~\bibnamefont{Sierou}} \bibnamefont{and}
  \bibinfo{author}{\bibfnamefont{J.~F.} \bibnamefont{Brady}},
  \bibinfo{journal}{J. Fluid Mech.} \textbf{\bibinfo{volume}{448}},
  \bibinfo{pages}{115} (\bibinfo{year}{2001}).

\bibitem[{\citenamefont{Bédorf et~al.}(2012)\citenamefont{Bédorf, Gaburov,
  and Zwart}}]{Bedorf:2012}
\bibinfo{author}{\bibfnamefont{J.}~\bibnamefont{Bédorf}},
  \bibinfo{author}{\bibfnamefont{E.}~\bibnamefont{Gaburov}}, \bibnamefont{and}
  \bibinfo{author}{\bibfnamefont{S.~P.} \bibnamefont{Zwart}},
  \bibinfo{journal}{J. Comp. Phys.} \textbf{\bibinfo{volume}{231}},
  \bibinfo{pages}{2825 } (\bibinfo{year}{2012}).

\bibitem[{\citenamefont{Bédorf and Zwart}(2012)}]{Bedorf:2012a}
\bibinfo{author}{\bibfnamefont{J.}~\bibnamefont{Bédorf}} \bibnamefont{and}
  \bibinfo{author}{\bibfnamefont{S.~P.} \bibnamefont{Zwart}},
  \bibinfo{journal}{Eur. Phys. J. Special Topics}  (\bibinfo{year}{2012}),
  \bibinfo{note}{in this issue}.

\end{thebibliography}

\end{document}